\documentclass[fleqn]{LMCS}
\input{jeffrey}
\usepackage{proof}
\usepackage{srcltx}
\usepackage{varioref}
\usepackage{mathpartir}
\usepackage{url}
\usepackage[latin1]{inputenc}
\usepackage[english]{babel}
\usepackage{amsfonts}
\usepackage{amsmath}
\usepackage{graphics}
\usepackage{enumerate,hyperref}
\newcommand{\gta}[1]{\goesto{#1}}
\newcommand{\rulee}[3]{\inferrule{#1}{#2}{~~\mathbf{#3}}}
\newcommand{\Gta}[1]{\Goesto{#1}}
\newcommand{\parop}{~|~}
\newcommand{\define}[0]{\stackrel{\mbox{{\tiny def}}}{=}}
\newcommand{\Longmapsto}{\mathopen{~|\mkern-9mu\Longrightarrow~}}
\newcommand{\nil}[0]{\mathbf{0}}
\newcommand{\tauPre}[1]{\tau.#1}
\newcommand{\inp}[3]{#1(#2).#3}
\newcommand{\out}[3]{\bar{#1}#2.#3}
\newcommand{\match}[3]{[#1=#2]#3}
\newcommand{\mismatch}[3]{[#1\neq #2]#3}
\newcommand{\ndc}[2]{#1+#2}
\newcommand{\para}[2]{#1\mathrel{|}#2}
\newcommand{\res}[2]{(\nu #1)#2}
\newcommand{\bang}[1]{!#1}

\newcommand{\weakCongSim}{\mathrel{\mathopen{\simeq\mkern-7mu >}}}
\newcommand{\weakCongSimOp}{\mathopen{\simeq\mkern-7mu >}}
\newcommand{\weakSimAux}{\mathopen{\approx\mkern-7mu >}}
\newcommand{\weakSimOp}[0]{\mathrel{\weakSimAux}}
\newcommand{\weakSim}[1]{\mathrel{\weakSimAux}_{#1}}
\newcommand{\fresh}{\mathrel{\sharp}}
\newcommand{\bisim}{\mathrel{\sim}}
\newcommand{\weakBisim}{\mathrel{\approx}}
\newcommand{\weakCong}{\mathrel{\cong}}

\newcommand{\picalc}[0]{$\pi$-calculus}
\newcommand{\new}[0]{$\ensuremath{\mathsf{N}}$}

\newcounter{ProofCounter}

\newcommand{\wli}[3]{\Goesto{#1 : #2  @ #3}}

\def\doi{5 (2:16) 2009}
\lmcsheading%
{\doi}
{1--36}
{}
{}
{Mar.~24, 2009}
{Jun.~30, 2009}
{}   

\begin{document}
\title{Formalising the $\pi$-calculus using Nominal Logic}
\author[J.~Bentson]{Jesper Bengtson}
\address{Department of Information Technology, University of Uppsala,
  Sweden}
\email{\{Jesper.Bengtson,Joachim.Parrow\}@it.uu.se}
\author[J.~Parrow]{Joachim Parrow}


\begin{abstract}
We formalise the pi-calculus using the nominal datatype package, based
on ideas from the nominal logic by Pitts et al., and demonstrate an
implementation in Isabelle/HOL. The purpose is to derive powerful
induction rules for the semantics in order to conduct machine
checkable proofs, closely following the intuitive arguments found in
manual proofs. In this way we have covered many of the standard
theorems of bisimulation equivalence and congruence, both late and
early, and both strong and weak in a uniform manner. We thus provide
one of the most extensive formalisations of a process calculus ever
done inside a theorem prover.

A significant gain in our formulation is that agents are identified up
to alpha-equivalence, thereby greatly reducing the arguments about
bound names. This is a normal strategy for manual proofs about the
pi-calculus, but that kind of hand waving has previously been
difficult to incorporate smoothly in an interactive theorem prover. We
show how the nominal logic formalism and its support in Isabelle
accomplishes this and thus significantly reduces the tedium of
conducting completely formal proofs. This improves on previous work
using weak higher order abstract syntax since we do not need extra
assumptions to filter out exotic terms and can keep all arguments
within a familiar first-order logic.
\end{abstract}

\maketitle

\section{Introduction}

\subsection{Motivation}
As the complexity of software systems increases, the need is growing
to ensure their correct operation. One way forward is to create
particular theories or frameworks geared towards particular
application areas. These frameworks have the right kind of
abstractions built in from the beginning, meaning that proofs can be
conducted at a high level. The drawback is that different areas need
different such frameworks, resulting in a proliferation and even
abundance of theories. A prime example can be found in the field of
process calculi. It originated in work by Milner in the late 1970s
\cite{MilCCS} and was intended to provide an abstract way to reason
about parallel and communicating processes. Today there are many
different strands of calculi addressing specific issues. Each of them
embodies a certain kind of abstraction suitable for a particular area
of application.

For each such calculus a certain amount of theoretical groundwork must
be laid down. Typical examples include definitions of the semantics,
establishing substitutive properties, structures for inductive proof
strategies etc. This groundwork must naturally be correct beyond doubt
(if there is an error in it then all proofs conducted in that calculus
will be incorrect).  The idea to use formal verification of the
groundwork itself is therefore natural. In this paper we shall present
an improved  method to accomplish this.

\subsection{Theorem provers}
 There exist today several proof assistants, aka theorem provers: Coq
 \cite{L:BC04}, Isabelle \cite{nipkow:isabelle}, Agda \cite{agda}, PVS
 \cite{cade92-pvs}, Nuprl \cite{nuprl-book} and HOL \cite{HOL}, just
 to name a few.  These theorem provers are interactive. They have many
 automated tactics, and the user can provide additional proof
 strategies.  Many are also getting better and easier to use, and so
 the concept of having fully machine checked proofs has recently
 become far more realistic. As an indication of this several major
 results have been proven over the last few years, including the four
 and five colour theorems \cite{bauer02colour, gonthier:fct}, Kepler's
 conjecture \cite{NipkowBS-IJCAR06} and G\"odel's incompleteness
 theorem~\cite{shankar:godel}. Significant advances in applications
 related to software are summarized in the POPLmark Challenge
 \cite{poplmark}, a set of benchmarks intended both for measuring
 progress and for stimulating discussion and collaboration in
 mechanizing the metatheory of programming languages. There are for
 example results on analysis of typing in system F and light versions
 of Java.  The theorem prover Isabelle is also currently used to
 verify software in the Verisoft project \cite{verisoft}.

We want to emphasize that these types of tools are now being
transferred to industry. In \cite{gordon:viisp}, a group at Microsoft
Research in Cambridge compiles a subset of $F\sharp$ (a Microsoft
product) code to the pi-calculus and security properties are checked
using ProVerif \cite{blanchet:proverif}. This work was later extendend
in \cite{bbfgm2008} where a cryptographic type checker was constructed
for $F\sharp$ which handles a larger set of problems. The ideas from
these are now being transferred into other Microsoft products. Also,
the Spec$\sharp$~\cite{specsharp} programming system is integrated in
the Microsoft Visual Studio environment for the .NET platform and
contains an automatic theorem prover.

\subsection{The $\pi$-calculus}
As the basic underlying model we have chosen the $\pi$-calculus, which
since its conception in the late 1980s ~by Milner, Parrow and
Walker~\cite{parrow:acomp} has had a significant impact on the way
formal methods handle mobile systems. The mechanism of name-passing,
in combination with the paradigm of static binding, where the scope of
names may be dynamically extended by means of communication to include
the receiver, has turned out to be surprisingly expressive for a vast
variety of programming idioms: abstract data types, lambda-calculus,
i.e.\ functional programming, object-oriented programming, imperative
programming, logic and concurrent constraint programming, and
primitives for encryption/decryption. The $\pi$-calculus has
influenced the development of many high-level programming languages
and it has triggered a whole family of related calculi. e.g.  spi
\cite{abadi97calculus}, join
\cite{FournetGonthier96:rcham-join-calculus}, fusion
\cite{victor:fusion-calculus}, blue \cite{boudol97picalculus}, the
applied $\pi$-calculus~\cite{abadi.fournet:mobile-values} and ambients
\cite{cardelli98mobile}. In essence, the $\pi$-calculus has now grown
out of a single formalism into a general field where components of
formalisms, such as operators, semantics and proof methods, can be
more freely combined.

\subsection{Approach}

The goal of our project is to provide a library in an automated
theorem prover, Isabelle/HOL \cite{nipkow:isabelle}, which allows
users to do machine checked proofs on the groundwork of process
calculi. The guiding principle is that the proofs should correspond
very closely to the traditional manual proofs present in the
literature. This means that for a person who has completed these
proofs manually very little extra effort should be required in order
to let Isabelle check them. Today those proofs are reasonably well
understood, but capturing them in a theorem prover has until now been
a daunting task. The reason is mainly related to bound names and the
desire to abstract away from $\alpha$-equivalence \cite{poplmark}.

In the literature it is not uncommon to find statements such as:
``henceforth we shall not distinguish between $\alpha$-equivalent
terms'' or ``we assume bound names to always be fresh'', even though
it is left unsaid exactly what this means. In
\cite{walkerSangiorgi:piCalculus} Sangiorgi and Walker write:
\begin{quote}
In any discussion, we assume that the bound names of any processes or actions under consideration are chosen to be different from the names free in any other entities under consideration, such as processes, actions, substitutions and sets of names.
\end{quote}
And in \cite{parrow:intro} we can find:
\begin{quote}
... we will use the phrase ``$\textit{bn}(\alpha)$ is fresh'' in a definition to mean that the name in $\textit{bn}(\alpha)$, if any, is different from any free name occurring in any of the agents in the definition.
\end{quote}

This kind of reasoning does not necessarily imply that proofs conducted in this manner are incorrect, only that they are not fully formalised.

Our approach is to formulate the $\pi$-calculus using ideas
from nominal logic developed by Pitts et al. \cite{PittsAM:nomlfo-jv,PittsAM:newaas-jv,PittsAM:nomu-jv}. This is a first order logic designed to work with
calculi using binders. It maintains all the properties of a first
order logic and introduces an explicit notion of freshness of names in the
terms.
Gabbay's thesis
\cite{gabbay:atoid}  uses it to introduce FM set theory, this is the
standard ZF set theory but with an extra axiom for freshness of
names.
Recent work by Urban and Tasson~\cite{urban:nti} extends this work using ideas from \cite{PittsAM:nomlfo-jv} and solves the problem with freshness without
introducing new axioms. The techniques have been implemented into the
theorem prover Isabelle/HOL, in a nominal datatype package, so that when defining nominal datatypes,
Isabelle will automatically generate a type which models the datatype
up to $\alpha$-equivalence as well as induction principles and a recursion combinator allowing the user to create functions on nominal datatypes.
\subsection{Results}
Our contribution is to use the nominal package in Isabelle to
describe the $\pi$-calculus. We have proved substantial portions of \cite{parrow:acomp} using these techniques. More specifically, we have proven that strong equivalence and weak congruence are congruence relations for both late and early operational semantics, that all structurally congruent terms are bisimilar and that late strong equivalence, weak bisimulation and weak congruence are included in their early counterparts. To our knowledge, properties about weak equivalences of the $\pi$-calculus have never before been formally derived inside a theorem prover. Our proof method is to lift the strong operational semantics to a weak one, enabling us to port our proofs between the two semantics. Moreover, our proofs follow their pen-and-paper equivalents very closely inside a first-order environment. In other words, the extra effort to have proofs checked by a machine is not prohibitive.
\subsection{Exposition}
In the next section we explain some basic concepts of the nominal
datatype package. We do not give a full account of it, only enough that a reader
may follow the rest of our paper. In Section \ref{kapOpsem} we cover the strong late operational semantics of the $\pi$-calculus as well as the induction and case analysis rules we have created for the semantic rules. Section \ref{kapBisim} treats strong late bisimulation, the proofs that it is preserved by all operators except input prefix and that strong equivalence is a congruence. In Section \ref{exampleDeriv} we show the proof strategies for one of our main results in depth demonstrating how closely our formalised proofs map their pen-and-paper equivalents. Section \ref{kapStructCong} handles the structural congruence rules and the proof that all structurally congruent terms are also bisimilar. We cover the weak late operational semantics in Section \ref{kapWeakBisim} and prove that weak bisimulation is preserved by all operators except sum and input prefix and that weak congruence is a congruence. In Section \ref{kapEarly} we formalise the early $\pi$-calculus, both strong and weak, and prove all the results which we have for the late semantics for early. We also prove that all late bisimulation relations are a subset of their corresponding early ones. In the concluding section we compare our efforts to related work and comment on planned further work. The Isabelle source files can be found at {\texttt http://www.it.uu.se/katalog/jesperb/pi}.

\section{The pi-calculus in Isabelle}

\noindent For a more thorough presentation of the nominal datatype
package in Isabelle the reader is referred to \cite{urban:nti}, but
enough basic definitions will be covered here for the reader to
understand the rest of this paper.  A {\em nominal datatype}
definition is like an ordinary data type but it explicitly tags the
binding occurrences of names. For example, a data type for
$\lambda$-calculus terms would in this way tag the name in the
abstraction. The point is that the nominal package in Isabelle
automatically generates induction rules where $\alpha$-equivalent
terms are identified, thus saving the user much tedium in large
proofs.

At the heart of nominal logic is the notion of {\em name swapping}
where names are a countably infinite set of atomic terms. If $T$ is
any term of permutation type (a term which supports permutations of
its names) and $a$ and $b$ are names then $(a~b)\bullet T$ denotes the
term where all instances of $a$ in $T$ become $b$ and vice versa.  All
names (even the binding and bound occurrences) are swapped in this
way. A {\em permutation} $p$ is a finite sequence of swappings. If $p
= (a_1~b_1)\cdots (a_n~b_n)$ then $p \bullet T$ means applying all
swappings in $p$ to $T$, beginning with the last element $(a_n~b_n)$.

Permutations are mathematically well behaved. They very rarely change
the properties of a term. Most importantly, $\alpha$-equivalence is
preserved by permutations. The property of
being preserved by permutations is often called
\emph{equivariance}. We shall mainly use equivariance on binary relations,
where the definition is:

\begin{defi}\label{equivariance}Equivariance
\[\mbox{\rm eqvt}~\mathcal{R}\define\forall p~T~U.~(T,~U)\in\mathcal{R}\Longrightarrow (p\bullet T,~p\bullet U)\in\mathcal{R}\]
\end{defi}

\noindent Another key concept is the notion of {\em support}. The definition, in
general, is that the support $\mbox{supp}~T$ of a term $T$ is the set
of names which can affect $T$ in permutations. In other words, if
 $p$ is a permutation only involving names outside the
support of $T$ then $p \bullet T = T$. Remembering that
$\alpha$-equivalent terms are identified we see that the support
corresponds to the {\em free names} in calculi like the
$\lambda$-calculus.

A crucial property is that the support of a term is finite. This
implies that for any term it is always possible to find a name outside
its support. We say that a name $a$ is {\em fresh} for a term $T$, written $a~\sharp~T$, if $a$ is not in the support of $T$.

Permutations can  be used to capture $\alpha$-equivalence. Let $[x].T$ stand for any operator that binds $x$ in $T$.
\begin{prop}\label{alphaeq}
$
 [x].T = [y].U\,\Longrightarrow\,
  (x = y \land T = U)\lor(x\neq y \land x~\sharp~U\land T=(x~y)\bullet U)
$
\end{prop}
\noindent If $[x].T = [y].U$ then either $x$ and $y$ are equal and $T$ and $U$ are $\alpha$-equivalent
or $x$ is not equal to $y$ and fresh in $U$ and $T$ is $\alpha$-equivalent to $U$
with all occurrences of $x$ swapped with $y$ and vice versa. Another way to capture $\alpha$-equivalence is the following:
\begin{prop}
\label{alphaeq2}
$
\begin{array}[t]{l}
 c~\sharp~(x, y, T, U) \land [x].T = [y].U\Longrightarrow (x~c)\bullet T=(y~c)\bullet U
\end{array}
$
\end{prop}
\noindent Here and
in the rest of the paper we use the word ``proposition'' for something
that Isabelle generates automatically.

We use a version of the monadic $\pi$-calculus
\cite{parrow:acomp}, and assume that the reader is familiar with the basic ideas
of its syntax and semantics.

\begin{defi}Defining the $\pi$-calculus in Isabelle.\vspace{5mm}
\\
\begin{tabular}{@{}rlc}
\multicolumn{2}{c}{{\bf Nominal declaration in Isabelle}}&{\bf Notation in this paper}\vspace{3mm}\\
{\tt nominal\_datatype pi =} & {\tt PiNil}& $\nil$\\
{\tt |}&{\tt Tau pi}&$\tauPre{P}$\\
{\tt |}&{\tt Input name "<<name>> pi}"&$\inp{a}{x}{P}$\\
{\tt |}&{\tt Output name name pi}&$\out{a}{b}{P}$\\
{\tt |}&{\tt Match name name pi}&$\match{a}{b}{P}$\\
{\tt |}&{\tt Mismatch name name pi}&$\mismatch{a}{b}{P}$\\
{\tt |}&{\tt Sum pi pi}&$\ndc{P}{Q}$\\
{\tt |}&{\tt Par pi pi}&$\para{P}{Q}$\\
{\tt |}&{\tt Res "<<name>> pi"}&$\res{x}{P}$\\
{\tt |}&{\tt Bang pi}&$\bang{P}$\\
\end{tabular}
\end{defi}
\noindent This definition is an example of Isabelle notation, where $\ll\mkern-9mu\mathit{name}\mkern-9mu\gg\mathit{pi}$ indicates that $\mathit{name}$ is bound in $\mathit{pi}$. For the rest of the paper we shall use the traditional notation for $\pi$-calculus terms as specified in the previous definition.

The nominal datatype package automatically generates lemmas for
reasoning about $\alpha$-equivalence between processes -- the ones
generated from Prop.\ \ref{alphaeq} can be found in the following proposition.
\begin{prop}The most commonly used $\alpha$-equivalence rules for the \emph{Input}- and the \emph{Restriction case}.\\

\begin{tabular}{ll}
Input:&$a(x).P = b(y).Q\Longrightarrow a=b \wedge
\begin{array}[t]{l}((x = y \wedge P = Q)~\vee\\(x\neq y \wedge x~\sharp~Q\wedge P=(x~y)\bullet Q))
\end{array}
$\\
Restriction:&$(\nu x)P = (\nu y)Q\Longrightarrow
\begin{array}[t]{l}(x = y \wedge P = Q)~\vee\\(x\neq y \wedge x~\sharp~Q\wedge P=(x~y)\bullet Q)
\end{array}
$
\end{tabular}
\label{alphaInputRes}
\end{prop}

\noindent Most modern theorem provers automatically generate induction rules for
defined datatypes. The nominal datatype package does the same for nominal
datatypes but with one addition: bound names which occur in the
inductive cases can be assumed to be disjoint from any finite set of
names. This greatly reduces the amount of manual
$\alpha$-conversions.

Functions over nominal datatypes have one restriction -- they may not
depend on the bound names in their arguments. Since nominal types are
equal up to $\alpha$-equivalence two equal terms may have different
bound names. When creating recursive functions over nominal datatypes
in Isabelle, one has to prove that this property holds for all
instantiations of the function. The nominal package provides the
appropriate proof conditions.

Our only nominal function is substitution where $P\{a/b\}$ (which can
be read $P$ with $a$ for $b$) is the agent obtained by replacing all
free occurrences of $b$ in $P$ with $a$.

\section{Operational semantics}
\label{kapOpsem}
\subsection{Definitions}
We use the standard operational semantics \cite{parrow:acomp}. Here
transitions are of the form $P \gta{\alpha} P'$, where $\alpha$ is an
action.  A first attempt, which works well for simpler calculi like
CCS, is to inductively define a set of tuples containing three
elements: a process $P$, an action $\alpha$ and the
$\alpha$-derivative of $P$ \cite{bengtson:gipc}.

However, in the $\pi$-calculus the action $\alpha$ may bind a name,
and the scope of this binding extends into $P'$. This observation is
made already in the original presentation of the $\pi$-calculus
\cite{parrow:acomp} where lemmas concerning variants of transitions
are spelled out. In his tutorial on the polyadic
pi-calculus~\cite{milner93polyadic} Milner uses "commitments" rather
than labelled transitions. A transition here corresponds to a pair
consisting of an agent and a commitment where the latter may have
binders and contains both the action and derivative process. We thus
face a discrepancy between a more traditional syntax for transitions
(looking like tuples of three elements) and the intended semantics
(that action and derivative in reality is one construct with names
that can be bound in all of it). In many presentations of the
$\pi$-calculus this issue is glossed over, and if $\alpha$-conversions
are not defined rigorously the three-element syntax for transitions
works fine. But here it poses a problem --- it would require us to
explicitly state the rules for changing the bound variable, and we
would not be able to rely on the otherwise smooth treatment of
$\alpha$-variants in our framework. Therefore, in our implementation
we follow \cite{milner93polyadic}, with a slight change of notation to
avoid confusion of prefixes and commitments, and define a
\emph{residual}-datatype which contains both action and derivative. It
binds the bound names of an action also in the derivative. (A similar
technique is also used by Gabbay when formalising the {\picalc} in FM
set theory \cite{GabbayMJ:picfm}.)
\begin{defi}The residual datatype
\label{residual}
\begin{verbatim}
datatype subject = Input name
                 | BoundOutput name

datatype freeRes = Output name name
                 | Tau
\end{verbatim}
\begin{verbatim}nominal_datatype residual = BoundResidual subject "<<name>> pi"
                          | FreeResidual freeRes pi
\end{verbatim}
\end{defi}

In this paper we shall continue to write pairs of processes and
residuals as transitions in the familiar way, and we need to
distinguish between actions that bind names and those that do not.  We
introduce the following notation.
\begin{defi}\hfil
\begin{enumerate}[(i)]
\item $P\gta{a\ll x\gg}P'$ denotes a transition with the bound name
  $x$ in the action. Note that $a$ is of type $\mathtt{subject}$. The
  residual by itself is written $a\mkern-5mu\ll\mkern-5mu
  x\mkern-5mu\gg~\prec P'$.

\item $P\gta{\alpha}P'$ denotes a transition without bound names. Note
  that $\alpha$ is of type $\mathtt{freeRes}$. The residual by itself
  is written $\alpha\prec P'$.

\item A transition can also be written as $P\longmapsto\mathit{Res}$
  where $P$ is an agent and $\mathit{Res}$ is a residual, for example
  $\tau.P\longmapsto\tau\prec P$
\end{enumerate}
\end{defi}

\noindent As previously mentioned, functions over nominal datatypes
cannot depend on bound names. This poses a slight problem, since
traditionally some of the operational rules have conditions on the
bound names. An example of this is the $\mathbf{Par}$ rule in the
standard operational semantics which states that the transition
$P\parop Q\gta{\alpha}P'\parop Q$ can occur only if ${P\gta{\alpha}P'}$
and $\mbox{bn}(\alpha)\cap\mbox{fn}(Q)=\emptyset$. A function such as
$\mbox{bn}$ does not exist in nominal logic and thus cannot be created
using the nominal datatype package.  An easy solution is to split the
operational rules which have these types of conditions into two rules
--- one for the transitions with bound names, and one for the ones
without. Doing this does not create extra proof obligations as most
proofs have to consider bound and free transitions separately anyway.

\noindent We can now define our operational semantics using inductively defined
sets which will contain pairs of processes and residuals. The Semantics, including the split
rules for {\bf Par} and {\bf Res} can be found in Fig.\ \ref{opSem}.


\begin{figure}
\fbox{\begin{minipage}{\hsize}
\begin{center}
\begin{mathpar}
\rulee{}{a(x).P\gta{a(x)}P}{Input}\and
\rulee{}{\bar{a}b.P\gta{\bar{a}b}.P}{Output}\and
\rulee{}{\tau.P\gta{\tau}P}{Tau}\and
\rulee{P\longmapsto\mathit{Res}}{[a=a]P\longmapsto\mathit{Res}}{Match}\and
\rulee{P\longmapsto\mathit{Res}~~~a\neq b}{[a\neq b]P\longmapsto\mathit{Res}}{Mismatch}\and
\rulee{P\gta{\bar{a}b}P'~~~a\neq b}{(\nu b)P\gta{\bar{a}(b)}P'}{Open}\and
\rulee{P\longmapsto\mathit{Res}}{P~+~Q\longmapsto\mathit{Res}}{Sum}\and
\rulee{P\gta{a\ll x\gg}P'~~~~x~\sharp~Q}{P~|~Q\gta{a\ll x\gg}P'~|~Q}{ParB}\and
\rulee{P\gta{\alpha}P'}{P~|~Q\gta{\alpha}P'~|~Q}{ParF}\and
\rulee{P\gta{a(x)}P'~~~Q\gta{\bar{a}b}Q'}{P~|~Q\gta{\tau}P'\{b/x\}~|~Q'}{Comm}\qquad\and
\rulee{P\gta{a(x)}P'~~~Q\gta{\bar{a}(y)}Q'~~~y~\sharp~P}{P~|~Q\gta{\tau}(\nu y)(P'\{y/x\}~|~Q')}{Close}\and
\rulee{P\gta{a\ll x\gg}P'~~~~y~\sharp~(a,~x)}{(\nu y)P\gta{a\ll x\gg}(\nu y)P'}{ResB}\and
\rulee{P\gta{\alpha}P'~~~~y~\sharp~\alpha}{(\nu y)P\gta{\alpha}(\nu y)P'}{ResF}\and
\rulee{P~|~!P\gta{\alpha}P'}{!P\gta{\alpha}P'}{Replication}
\end{mathpar}
\end{center}
\caption{The \emph{Par}- and the \emph{Res}-rule in the operational semantics of the $\pi$-calculus have been split. Symmetric versions have been elided.}
\label{opSem}
\end{minipage}}
\end{figure}

As mentioned previously, permutations are usually very well behaved. The following proposition is generated automatically by the nominal package.
\begin{prop}$P\longmapsto\mbox{Res}\Longrightarrow p\; \mathrel{\bullet}P\longmapsto p\; \bullet\mbox{Res}$
\label{semEqvt}
\end{prop}
\newpage
\subsection{Induction and case analysis rules}
\subsubsection{Automatically generated rules}
Isabelle will automatically create rules for both induction and case
analysis of the semantics. They are specifically tailored to allow
induction over all possible transitions but can also be custom made to
do induction or case analysis over specific types of processes, such
as those composed by the $|$-operator. They will have an assumption of
the form $P\longmapsto\mathit{Res}$, which is the term with which we
are working, and a logical proposition $\mathit{Prop}$ which is what
we want to prove. When applied these rules will generate a set of
subgoals where every subgoal corresponds to one action that the
process $P$ could take to end up in $\mathit{Res}$ -- in short,
$\mathit{Prop}$ needs to be proven for all possible transitions for
the rule to hold. The rules do, however, assume that the equivalence
relation used is syntactic equivalence and not
$\alpha$-equivalence. The nominal datatype package automatically
creates induction rules for nominal datatypes as well as for
inductively defined sets or predicates. The induction rule generated
for the semantics is the largest possible one which does induction
over all operational rules and there is currently no way to
automatically generate case analysis rules for transitions of a
certain form. To derive rules for the cases which do not make use of
bound names is unproblematic. In fact, Isabelle will be able to derive
the following case analysis rules with very little help.
\begin{prop}
\label{tauCases}
The automatically generated case analysis rule for tau-transitions.\\
\[
\infer{\mbox{Prop}}{
   \deduce{\alpha = \tau \land P = P'\Longrightarrow\mbox{Prop}}{
       \tau.P\gta{\alpha}P'\\
   }
}
\]
\end{prop}
\begin{prop}
\label{outputCases}
The automatically generated case analysis rule for output transitions.\\
\[
\infer{\mbox{Prop}}{
   \deduce{\alpha = \bar{a}b \land P = P'\Longrightarrow\mbox{Prop}}{
      \bar{a}b.P\gta{\alpha}P'
    }
}
\]
\end{prop}
\begin{prop}
\label{matchCases}
The automatically generated case analysis rule for matches.\\
\[
\infer{\mbox{Prop}}{
   \deduce{a=b\land P\longmapsto\mbox{Res}\Longrightarrow\mbox{Prop}}{
      [a=b]P\longmapsto\mbox{Res}
   }
}
\]
\end{prop}
\begin{prop}
\label{mismatchCases}
The automatically generated case analysis rule for mismatches.\\
\[
\infer{\mbox{Prop}}{
   \deduce{a \neq b\land P\longmapsto\mbox{Res}\Longrightarrow\mbox{Prop}}{
      [a\neq b]P\longmapsto\mbox{Res}
   }
}
\]
\end{prop}

\begin{prop}
\label{sumCases}
The automatically generated case analysis rules for sums.\\
\[
\infer{\mbox{Prop}}{
   \deduce{
           \begin{array}{l}
           P\longmapsto\mbox{Res}\Longrightarrow\mbox{Prop}\\
           Q\longmapsto\mbox{Res}\Longrightarrow\mbox{Prop}
           \end{array}
          }{
      P + Q\longmapsto\mbox{Res}
   }
}
\]
\end{prop}
\noindent The rest of the rules generated by Isabelle for our operational semantics deal with bound names and suffer from three problems, which we now address in turn.
\subsubsection{Problems with generated bound names}
The first problem is that some semantic case analysis rules generate bound names. When the
rule is applied in the context of a proof, there is no a priori
guarantee that these names are fresh in this larger context. We
therefore  derive rules for induction and case analysis which are parameterized on a
finite set of names, the ``context names'', which the user can provide when applying the rule. The bound names generated by the rules are guaranteed to
be fresh from the context names (just as is guaranteed
for induction rules genereated by the nominal package, and for the same reason: avoiding name clashes
and $\alpha$-conversions later in the proof). This idea stems from \cite{urban:nti} but was developed independently of similar work in \cite{urban:barendregt}. The logical framework has also been covered in \cite{PittsAM:alpsri}.

As an example a derived rule for case analysis of the parallel
operator is shown in the following proposition where the parameter $\mathcal{C}$ represents a set of context names and can be instantiated with any nominal datatype:
\begin{lem}The derived case analysis rule for the parallel operator with no bound names in the transition.\\
\label{parCasesF}
\[
\infer{\mbox{Prop}}{
   \deduce{
      \begin{array}{l}
      \forall P'.~P\gta{\alpha}P'\land R = P'~|~Q\Longrightarrow \mbox{Prop}\\
      \forall Q'.~Q\gta{\alpha}Q'\land R = P~|~Q'\Longrightarrow \mbox{Prop}\\
      \begin{array}{@{}l@{\; }l}
      \forall P'~Q'~a~x~b.&P\gta{a(x)}P'\land Q\gta{\bar{a}b}Q'\land\alpha=\tau\land\\
      ~&R = P'\{b/x\}~|~Q'\land x~\sharp~\mathcal{C}\Longrightarrow\mbox{Prop}\\
      \forall P'~Q'~a~x~b.&P\gta{\bar{a}b}P'\land Q\gta{a(x)}Q'\land\alpha=\tau\land\\
      ~&R = P'\parop Q'\{b/x\}\land x~\sharp~\mathcal{C}\Longrightarrow\mbox{Prop}
      \end{array}\\
      \begin{array}{@{}l@{\; }l}
      \forall P'~Q'~a~x~y.&P\gta{a(x)}P'\land Q\gta{\bar{a}(y)}Q'\land y~\sharp~P\land\alpha=\tau~\land\\
      ~&R = (\nu y)(P'\{y/x\}~|~Q')\land x~\sharp~\mathcal{C}\land y~\sharp~\mathcal{C}\Longrightarrow\mbox{Prop}\\
      \forall P'~Q'~a~x~y.&P\gta{\bar{a}(y)}P'\land Q\gta{a(x)}Q'\land y~\sharp~Q\land\alpha=\tau~\land\\
      ~&R = (\nu y)(P'\parop Q'\{y/x\})\land x~\sharp~\mathcal{C}\land y~\sharp~\mathcal{C}\Longrightarrow\mbox{Prop}
      \end{array}
      \end{array}
     }{
     P~|~Q\gta{\alpha}R
   }
}
\]
\label{parStruct}
\end{lem}
\noindent Each all-quantified term corresponds to a possible transition by the process $P\parop Q$. The two semantic rules which
introduce bound names are the \emph{Comm}- and the \emph{Close}
rules. The rule can be instantiated with an arbitrary term $\mathcal{C}$ and these bound names will be set fresh for
that term.
\subsubsection{Problems with equivalence checks on terms}
The second problem is that in
case analysis, equivalence checks between terms always appear. If
these  terms contain bound names, such as $(\nu x)P = (\nu y)Q$, then
normal unification is not possible. As seen in
Prop.\ \ref{alphaeq} and \ref{alphaeq2}, every such equivalence check produces either two
cases which both have to be proven or one case with several permutation and freshness conditions.
As an example, a rule for
case analysis on the $\nu$-operator with no bound names in the
action can be found in the following proposition:
\begin{prop}\label{nuStruct} The automatically generated case analysis rule for the $\nu$-operator, based on Prop.\ \ref{alphaeq}, where no bound name occurs in the action.
\[
\infer{\mbox{Prop}}{
   \deduce{\forall Q~Q'~\beta~y.~Q\gta{\beta}Q'\land y~\sharp~\beta\land(\nu x)P = (\nu y)Q\land\alpha=\beta\land P'=(\nu y)Q'\Longrightarrow\mbox{Prop}}{
        (\nu x)P\gta{\alpha}P'
   }
}
\]
\end{prop}

\noindent The conjunct $(\nu x)P = (\nu y)Q$
poses a problem as we have to show {\em Prop} for all cases such
that the equivalence holds. We can reason about this equality using either Prop.\ \ref{alphaeq} or Prop.\ \ref{alphaeq2} but neither of these rules are convenient to work with. Prop.\ \ref{alphaeq} causes a case explosion which forces us to prove the same thing several times for different permutations on terms and Prop.\ \ref{alphaeq2} introduces extra permutations which makes the proof more cumbersome to work with.
We therefore use the following derived lemma in place of the original case analysis rule:

\begin{lem}\label{resStructInduct}
Case analysis rule derived from Prop.\ \ref{nuStruct}.
\[
\infer{\mbox{Prop}}{
   \deduce{\forall P''.~P\gta{\alpha}P''\land x~\sharp~\alpha\land P'=(\nu x)P''\Longrightarrow\mbox{Prop}}{
      (\nu x)P\gta{\alpha}P'
   }
}
\]
\end{lem}
\noindent The main idea of the proof is to find a $P''$ which suitably
depends on the universally quantified terms in the second assumption
of the original proposition.

The other rule which require this treatment is the case analysis rule for the parallel operator where the transition contains a bound name.
\begin{lem}\label{parCasesB}
Case analysis rule for the parallel operator with a bound name in the
transition. 
\[
\infer{\mbox{Prop}}{
   \deduce{
       \begin{array}{l}
       \forall P'.~P\gta{a\ll x\gg}P'\land x\fresh Q\land R = P'\parop Q\Longrightarrow\mbox{Prop}\\
       \forall Q'.~Q\gta{a\ll x\gg}Q'\land x\fresh P\land R = P\parop Q'\Longrightarrow\mbox{Prop}
       \end{array}
    }{
        P\parop Q\gta{a\ll x\gg} R\\
    }
}
\]
\end{lem}

\subsubsection{Problems with multiple bound names in terms}
The third problem arises when several bound names occur in the term that you want to do case analysis on. We have already shown how we can ensure that any newly generated bound names are disjoint from any context we might be interested in. The problem here is that since multiple bound names are present before case analysis starts, any properties regarding them are fixed in the environment and if we have a name clash, we have to do manual $\alpha$-conversions. There are two rules that suffer from this problem, where the simplest one is the one for input-prefix. To solve this problem, we derive the following case analysis rule.
\begin{lem}\label{inputCases}
The derived case analysis rule for the input-prefix.
\[
\infer{\mbox{Prop}}{
   \deduce{a = b\land P' = (x~y)\bullet P\Longrightarrow\mbox{Prop}}{
      a(x).P\gta{b(y)}P'
   }
}
\]
\end{lem}

\noindent The other rule which requires this treatment is the restriction case where a bound name appears in the transition.
\begin{lem}
\label{resCasesB}
The derived case analysis rule for restriction with a bound name in the transition.\\
\[
\infer{\mbox{Prop}}{
   \deduce{
      \begin{array}{l}
      \forall b~P''.~P\gta{\bar{b}x}P''\land b\neq x\land a = \mathtt{BoundOutput}~b\land P' = (x~y)\bullet P''\Longrightarrow\mbox{Prop}\\
      \forall P''.~P\gta{a\ll y\gg}P''\land x\fresh a\land x\neq y\land P' = (\nu x)P''\Longrightarrow\mbox{Prop}
      \end{array}
    }{
      \deduce{x\neq y}{(\nu x)P\gta{a\ll y\gg}P'}
    }
}
\]
\end{lem}
\noindent In this rule we require $x$ and $y$ to be disjoint. The two applicable rules from the semantics have conflicting requirements on the bound names -- one requires them to be the same, and the other requires them to be disjoint. To keep the generality of the lemma, we keep the bound names disjoint and in the $\mathbf{Open}$ case permute the names in the derivative. As we shall se later, we will always be in a context where we can guarantee that $x$ and $y$ are separate when applying this rule.
\subsubsection{Induction}
The remaining operator is the $!$-operator which requires an induction rule rather than a case analysis rule as it is the only operator which occurs in the premise of its inference rule, as can be seen in Fig. \ref{opSem}. As in Lemma \ref{parCasesF}, $\mathcal{C}$ is a parameter representing the names with which new bound names may not clash.
\begin{lem}
\label{bangCases}
The derived induction rule for the $!$-operator.\\
\begin{center}
\infer{\mbox{Prop}~\mathcal{C}~(!P)~\mbox{Res}}{
   \deduce{
      \begin{array}{l}
      (1)~\forall a~x~P'~\mathcal{C}.~P\gta{a\ll x\gg}P'\land x\fresh P\land x\fresh\mathcal{C}\Longrightarrow\mbox{Prop}~\mathcal{C}~(P\parop!P)~(a\mkern-5mu \ll\mkern-5mu x\mkern-5mu\gg~\prec P'\parop!P)\\
      (2)~\forall\alpha~P'~\mathcal{C}.~P\gta{\alpha}P'\Longrightarrow\mbox{Prop}~\mathcal{C}~(P\parop!P)~(\alpha\prec P'\parop!P)\\
      \begin{array}{@{}l@{\; }l}
      (3)~\forall a~x~P'~\mathcal{C}.&!P\gta{a\ll x\gg}P'\land x\fresh P\land x\fresh\mathcal{C}\land\mbox{Prop}~\mathcal{C}~(!P)~(a\mkern-5mu\ll\mkern-5mu x\mkern-5mu\gg~\prec P')\Longrightarrow\\
      ~&\mbox{Prop}~\mathcal{C}~(P\parop!P)~(a\mkern-5mu\ll\mkern-5mu x\mkern-5mu \gg~\prec P\parop P')\\
      \end{array}\\
      (4)~\forall\alpha~P'~\mathcal{C}.~!P\gta{\alpha}P'\land\mbox{Prop}~\mathcal{C}~(!P)~(\alpha\prec P')\Longrightarrow\mbox{Prop}~\mathcal{C}~(P\parop!P)~(\alpha\prec P\parop P'')\\
      \begin{array}{@{}l@{\; }l}
      (5)~\forall a~x~b~P'~P''~\mathcal{C}.&P\gta{a(x)}P'\land~!P\gta{\bar{a}b}P''\land\mbox{Prop}~\mathcal{C}~(!P)~(\bar{a}b\prec P'')\land x\fresh\mathcal{C}\Longrightarrow\\
      ~&\mbox{Prop}~\mathcal{C}~(P\parop!P)~(\tau\prec P'\{b/x\}\parop P'')\\
      \end{array}\\
      \begin{array}{@{}ll}
      (6)~\forall a~x~b~P'~P''~\mathcal{C}.&P\gta{\bar{a}b}P'\land~!P\gta{a(x)}P''\land\mbox{Prop}~\mathcal{C}~(!P)~(a(x)\prec P'')\land x\fresh\mathcal{C}\Longrightarrow\\
      ~&\mbox{Prop}~\mathcal{C}~(P\parop!P)~(\tau\prec P'\parop P''\{b/x\})\\
      \end{array}\\
      \begin{array}{@{}l@{\; }l}
      (7)~\forall a~x~y~P'~P''~\mathcal{C}.&P\gta{a(x)}P'\land~!P\gta{\bar{a}(y)}P''\land\mbox{Prop}~\mathcal{C}~(!P)~(\bar{a}(y)\prec P'')\land x\fresh\mathcal{C}~\land\\
      ~&y\fresh P\land y\fresh\mathcal{C}\Longrightarrow\mbox{Prop}~\mathcal{C}~(P\parop!P)~(\tau\prec(\nu y)(P'\{y/x\}\parop P''))\\
      \end{array}\\
      \begin{array}{@{}l@{\; }l}
      (8)~\forall a~x~y~P'~P''~\mathcal{C}.&P\gta{\bar{a}(y)}P'\land~!P\gta{a(x)}P''\land\mbox{Prop}~\mathcal{C}~(!P)~(a(x)\prec P'')\land x\fresh\mathcal{C}~\land\\
      ~&y\fresh P\land y\fresh\mathcal{C}\Longrightarrow\mbox{Prop}~\mathcal{C}~(P\parop!P)~(\tau\prec(\nu y)(P'\parop P''\{y/x\}))\\
      \end{array}
      \end{array}
   }{
   !P\longmapsto\mbox{Res}
   }
}
\end{center}
\end{lem}
\noindent Each numbered line corresponds to one way that an action can be inferred from a replication. Line (1) and (2) cover the case where a single process makes an action, line (3) and (4) perform the inductive step where a process in the smaller chain of replicated processes makes an action. Line (5) and (6) handle communication and line (7) and (8) handle scope extrusion.

The derived lemma is an induction rule in that it has the induction hypothesis $\emph{Prop}$ occurring on the left hand side of the implications in the inductive rules where $!$ occurs. A simpler rule which only makes use of the inference rule for ! is available, but the proofs we are interested in would have to make use of the rules for the $|$-operator to reason about all possible transitions that a process of the form $!P$ could do. This induction rule combines the two in one rule.

\section{Strong bisimulation}
\label{kapBisim}
\subsection{Simulation}
 Intuitively, two processes are said to be bisimilar if they can
 mimic each other step by step. Traditionally, a bisimulation is a
symmetric binary relation $\mathcal{R}$ such that for all processes
$P$ and $Q$ in $\mathcal{R}$, if $P$ can do an action, then $Q$ can
mimic that action and their corresponding derivatives are in
$\mathcal{R}$.

When defining bisimulation between two processes in the
$\pi$-calculus, extra care has to be taken with respect to bound names
in actions. Consider the following processes:
\[\begin{array}{l}
    P \define a(u).(\nu b)\bar{b}x.0\\
    Q \define a(x).0
  \end{array}
\]
Clearly $P$ and $Q$ should be bisimilar since they both can do only
one input action along a channel $a$ and then nothing more. But since
$x$ occurs free in $P$, $P$ cannot be $\alpha$-converted into
$a(x).(\nu b)\bar{b}x$. However, since processes have finite support,
there exists a name $w$ which is fresh in both $P$ and $Q$ and after
$\alpha$-converting both processes, bisimulation is possible. Hence,
when reasoning about bisimulation, we must restrict attention to
the bound names of actions which are fresh for both $P$ and $Q$. One
of our main contributions is how this is achieved without running into
a multitude of $\alpha$-conversions.
Our formal definition of bisimulation equivalence uses the following notion, where $\mathcal{R}$ is a binary relation on agents.

\begin{defi}
\label{simulation}
The agent $P$ can simulate the agent $Q$ preserving $\mathcal{R}$, written $P\leadsto_\mathcal{R}Q$, if
\[
\begin{array}{@{}ll}
(\forall a~x~Q'.&Q\gta{a\ll x\gg}Q' \land x~\sharp~P \Longrightarrow\\
~&\exists P'.~P\gta{a\ll x\gg}P' \land \mbox{\rm derivative}(a,~x,~P',~Q',~\mathcal{R}))~\land
\end{array}\\
\]
\[
(\forall \alpha~Q'.~Q\gta{\alpha}Q' \Longrightarrow \exists P'.~P\gta{\alpha}P' \land (P', Q') \in \mathcal{R})
\]
  where
\[
\mbox{\rm derivative}(a,~x,~P',~Q',~\mathcal{R})\define
\]
\[
\begin{array}{rlll}
\mathbf{case}~a~\mathbf{of}&\mbox{\rm Input}~\_&\Rightarrow&\forall u.~(P'\{u/x\},~Q'\{u/x\})\in\mathcal{R}\\
|&\mbox{\rm BoundOutput}~\_&\Rightarrow &(P',~Q')\in\mathcal{R}
\end{array}
\]
\end{defi}

\noindent Note that the argument $a$ in $\mbox{\rm derivative}$ is of type
\texttt{subject} as described in Def. \ref{residual}.  Thus, the
requirement is that if $Q$ has an
action then $P$ has the same action, and the derivatives $P'$ and $Q'$
are in $\mathcal{R}$. 

Equivariance also needs to be established for simulations. More specifically, we need to prove the following lemma:
\begin{lem} If $P\leadsto_\mathcal{R} Q$, $\mathcal{R}$ is a subset of $\mathcal{R'}$ and $\mathcal{R'}$ is equivariant then $p \bullet P\leadsto_\mathcal{R'}p\bullet Q$.
\label{simEqvt}
\begin{proof}
By Def. \ref{simulation}. The intuition is to apply the inverse permutation of $p$ to cancel it out. The inverse can be applied using Lemma \ref{semEqvt} and the assumption $\mbox{\rm eqvt}~\mathcal{R'}$.
\end{proof}
\end{lem}

\noindent The traditional way to define strong bisimulation equivalence is to
say that $\mathcal{R}$ is a {\em bisimulation} if it is symmetric and
that for all agents $P, Q$ it holds that $(P,Q)\in \mathcal{R}
\to P\leadsto_\mathcal{R}Q$; the strong bisimulation
equivalence is then the union of all strong bisimulations. As we shall
see in a moment, an alternative definition using direct coinduction, similar to the approach in \cite{honsell:picalculus},
yields shorter proofs. Our main improvement, however, is in the
treatment of the bound name $x$. In Def. \ref{simulation} it is by definition ensured not to be
among the free names in $P$, but when we use it within a complex proof
we will run into a massive case analysis on whether $x$ is equal to
other names used in the proof. In the same way as in Lemma \ref{parCasesF} we
bypass this tedium and derive the following introduction
rule for an arbitrary nominal data term $\mathcal{C}$. This term is provided by the user to ensure that the bound name is distinct from any name occurring so far in the proof.

\begin{lem}An introduction rule for simulation avoiding name clashes.\\
\label{simintro}
\[
\infer{P\leadsto_\mathcal{R}Q}{
  \deduce{
      \begin{array}{@{}l@{\; }l}
      \forall a~x~Q'.&Q\gta{a\ll x\gg}Q' \land x~\sharp~\mathcal{C} 
                         \Longrightarrow\\
      ~&\exists P'.~P\gta{a\ll x\gg}P' \land 
                    \mbox{\rm derivative}(a,~x,~P',~Q',~\mathcal{R})\\
      \multicolumn{2}{@{}l}{\forall \alpha~Q'.~Q\gta{\alpha}Q' \Longrightarrow \exists P'.~P\gta{\alpha}P' \land (P', Q') \in \mathcal{R}}
      \end{array}
      }{
      \mbox{\rm eqvt}~\mathcal{R}
   }
}
\]
\end{lem}
\noindent This is used extensively in our proofs. We can in this way make
sure that whenever bound names appear in our proof context, these
bound names do not clash with other names which would force us to do
$\alpha$-conversions. The amount of $\alpha$-conversions we have to do
manually is reduced to the instances where they would be required in a manual proof.

Note that we need an extra requirement that our simulation relation
is equivariant. The reason is that if the relation is not closed under
permutations, we cannot $\alpha$-convert our processes. Fortunately,
all relations of interest turn out to be equivariant and the proofs trivial.

\subsection{Preservation properties}
Our simulations are parametrised on an arbitrary relation $\mathcal{R}$. We exploit this by providing, for each operator, a set of constraints on $\mathcal{R}$ such that the operator preserves $\leadsto_\mathcal{R}$. This set of constraints should be kept as small as possible as they will have to be proven when we prove preservation properties of bisimulation. In this section we show all proofs that are needed to show that a relation is preserved by all operators.

We first establish lemmas for reflexivity and transitivity.

\begin{lem}
\label{simRefl}
$\mbox{\rm Id}\subseteq\mathcal{R}\Longrightarrow P\leadsto_\mathcal{R}P$
\end{lem}
\begin{proof}
By the definition of simulation.
\end{proof}
\begin{lem}
\label{simTrans}
$P\leadsto_\mathcal{R} Q\land Q\leadsto_{\mathcal{R}'} R\land\mbox{\rm eqvt}~\mathcal{R''}\land\mathcal{R'}~\circ~\mathcal{R}\subseteq\mathcal{R''}\Longrightarrow P\leadsto_{\mathcal{R''}} R$
\end{lem}
\begin{proof}
By Lemma \ref{simintro} and setting $\mathcal{C}$ to $(P,~Q)$ to make the bound names which occur in the transitions disjoint from $P$ and $Q$. We would otherwise have to do manual $\alpha$-conversions when traversing the simulation chain.
\end{proof}
\noindent We can now move on to our preservation lemmas.
\begin{lem}
\label{tauPres}
$(P,~Q)\in\mathcal{R}\Longrightarrow\tau.P\leadsto_\mathcal{R}\tau.Q$
\end{lem}
\begin{proof}
By Def.\ \ref{simulation} and Prop.\ \ref{tauCases}.
\end{proof}
\begin{lem}
\label{outputPres}
$(P,~Q)\in\mathcal{R}\Longrightarrow\bar{a}b.P\leadsto_\mathcal{R}\bar{a}b.Q$
\end{lem}
\begin{proof}
By Def.\ \ref{simulation} and Prop.\ \ref{outputCases}.
\end{proof}
\noindent In order for a relation to preserved by the input prefix it needs to be closed under substitutions. We write $\mathcal{R}^s$ for the closure of the relation $\mathcal{R}$ under all substitutions.
\begin{defi}
  $P~\mathcal{R}^s~Q\define\forall\sigma.~(P\sigma)\mathrel{\mathcal{R}}(Q\sigma)$ where $\sigma$ is a chain of substitutions.
\end{defi}
\begin{lem}
\label{inputPres}
$(P,~Q)\in\mathcal{R}^s\land\mbox{\rm eqvt}~\mathcal{R}\Longrightarrow a(x).P\leadsto_\mathcal{R}a(x).Q$
\end{lem}
\begin{proof}
By Lemma \ref{simintro} and setting $\mathcal{C}$ to $(x,~P)$. Lemma \ref{inputCases} can then be used to finish the proof.
\end{proof}
\begin{lem}
\label{matchPres}
\[
\infer{[a=b]P\leadsto_\mathcal{R}[a=b]Q}{
   P\leadsto_\mathcal{R}Q
}
\]
\end{lem}
\begin{proof}
By Def.\ \ref{simulation} and Prop.\ \ref{matchCases}.
\end{proof}
\begin{lem}
\label{mismatchPres}
\[
\infer{[a\neq b]P\leadsto_\mathcal{R}[a\neq b]Q}{
   P\leadsto_\mathcal{R}Q
}
\]
\end{lem}
\begin{proof}
By Def.\ \ref{simulation} and Prop.\ \ref{mismatchCases}.
\end{proof}
\begin{lem}
\label{sumPres}
\[
\infer{P+S\leadsto_\mathcal{R}Q+S}{
   \deduce{\mbox{\rm Id}\subseteq\mathcal{R}}{P\leadsto_\mathcal{R}Q}
}
\]
\end{lem}
\begin{proof}
By Def.\ \ref{simulation}, Prop. \ \ref{sumCases} and Lemma \ref{simRefl}.
\end{proof}
\noindent The remaining preservation lemmas do not require that the relation reasoned about in the assumptions are the same as in the conclusions. It suffices to require them to be related by a set of constraints. The reason for this will be clarified when we cover bisimulation, suffice here to say that it makes the lemmas more general.
\begin{lem}~\\
\label{parPres}
\[
\rulee{
    P\leadsto_\mathcal{R} Q \and
    (P,~Q)\in\mathcal{R} \and
    \mbox{\rm Id}\subseteq\mathcal{R} \\\\\
    \forall P~Q~S.~(P,~Q)\in\mathcal{R}\Longrightarrow(P\parop S,~Q\parop S)\in\mathcal{R'} \\\\
    \forall P~Q~x.~(P,~Q)\in\mathcal{R'}\Longrightarrow((\nu x)P,~(\nu x)Q)\in\mathcal{R'}
}
{P\parop S\leadsto_\mathcal{R'}Q\parop S}{}
\]
\end{lem}
\begin{proof}
By the definition of $\leadsto$. Lemma \ref{parCasesB} is used to prove the cases where bound names occur in the transition and Lemma \ref{parCasesF} is used otherwise. When using Lemma \ref{parCasesF} $\mathcal{C}$ is set to $(P,~S)$. This proof will be covered more extensively in Section \ref{exampleDeriv}.
\end{proof}
\begin{lem}~\\
\label{resPres}
\[
\rulee{
P\leadsto_\mathcal{R} Q \and
\mbox{\rm eqvt}~\mathcal{R} \and
\mbox{\rm eqvt}~\mathcal{R'} \and
\mathcal{R}\subseteq\mathcal{R'} \\\\
\forall P~Q~x.~(P,~Q)\in\mathcal{R}\Longrightarrow((\nu x)P,~(\nu x)Q)\in\mathcal{R'}
}
{
   (\nu x)P\leadsto_\mathcal{R'}(\nu x)Q
}{}
\]
\end{lem}
\begin{proof}
By Lemma \ref{simintro} and setting $\mathcal{C}$ to $(x,~P)$. Lemma \ref{resCasesB} and \ref{nuStruct} are then used for the case analysis. $\mathcal{R}$ has to be equivariant since the $\mathbf{Open}$ case introduces permutations which need to be applied to the relation.
\end{proof}
\noindent The remaining preservation lemma we need is for the $!$-operator. For this proof we are going to need a recursively defined relation. This follows from the fact that the $!$-operator is the only operator which occurs on the left hand side of the semantic rules and the proof needs to be done on the depth of inference and not the size of the term.
\begin{defi}~\\
\[
\begin{array}{l@{\; }l}
\mbox{\rm Rep}~\mathcal{R}\define&(P,~Q)\in\mathcal{R}\Longrightarrow(!P,~!Q)\in\mbox{\rm Rep}~\mathcal{R}\\
~&(P,~Q)\in\mathcal{R}\land(S,~T)\in\mbox{\rm Rep}~\mathcal{R}\Longrightarrow(P\parop S,~Q\parop T)\in\mbox{\rm Rep}~\mathcal{R}\\
~&(P,~Q)\in\mbox{\rm Rep}~\mathcal{R}\Longrightarrow((\nu x)P,~(\nu x)Q)\in\mbox{\rm Rep}~\mathcal{R}
\end{array}
\]
\end{defi}
\begin{lem}~\\
\label{bangPres}
\[
\rulee{
(P,~Q)\in\mathcal{R} \and
\mbox{\rm eqvt}~\mathcal{R} \\\\
\forall P~Q.~(P,~Q)\in\mathcal{R}\Longrightarrow P\leadsto_\mathcal{R} Q
}
{
   !P\leadsto_{\mbox{\rm Rep}~\mathcal{R}} !Q
}{}
\]
\end{lem}
\begin{proof}
The trick here is to include the fact that $(!P,~!Q)\in\mbox{\rm Rep}~\mathcal{R}$ in the induction hypothesis. We use Lemma \ref{bangCases} to do induction over the transitions made by the process $!Q$. We know that the processes in the relation $\mathcal{R}$ simulate each other and the induction hypothesis generates the simulations by the nested replications. This proof is the most extensive of the preservation proofs due to its many cases and the need for induction.
\end{proof}
\subsection{Strong Bisimulation}
Strong bisimulation equivalence can be described using coinduction, i.e. the
greatest fixed point derived from a monotonic function.

\begin{defi}Strong bisimulation equivalence, $\sim$, is the largest relation satisfying:
\label{bisimulation}
\[
  P \sim Q \Longrightarrow P \leadsto_\sim Q~\land~Q\leadsto_\sim P
\]
\end{defi}

\noindent Note that we do not need to define what a bisimulation is; our
coinductive definition uses $P\leadsto_\mathcal{R}Q$
directly. This defines $\sim$ to be the largest relation such that related
agents can simulate each other preserving $\sim$.

Conducting proofs on bisimulation equivalence often boils
down to proving the same thing twice -- once for each direction. With
our formulation it is often easy to just prove one direction and let
the other be inferred automatically.

When proving that two processes are bisimilar, we pick a
set $\mathcal{X}$ which contains the processes and which
respects the constraints of the corresponding preservation lemma. It then suffices to show
that all members of $\mathcal{X}$ are simulated preserving
$\mathcal{X}~\cup \sim$. The following coinduction rules are easily derivable from the ones genereated by Isabelle.
\begin{lem}
\label{bisimCoinduct}
\[
\rulee{
(P,~Q)\in\mathcal{X} \\\\
\forall P'~Q'.~(P',~Q')\in\mathcal{X}\Longrightarrow P'\leadsto_\mathcal{X\cup\sim} Q'\land Q'\leadsto_\mathcal{X\cup\sim} P'
}
{
   P\sim Q
}{}
\]
\end{lem}
\begin{lem}
\label{bisimWeakCoinduct}
\[
\rulee{
(P,~Q)\in\mathcal{X} \\\\
\forall P'~Q'.~(P',~Q')\in\mathcal{X}\Longrightarrow P'\leadsto_\mathcal{X} Q'\land Q'\leadsto_\mathcal{X} P'
}
{
  P \sim Q
}{}
\]
\end{lem}
\noindent The difference between the two rules is found in the goal where the weaker version requires the processes to be simulated preserving $\mathcal{X}\; \cup\sim$ whereas the stronger version only requires them to be simulated preserving $\mathcal{X}$. Unless otherwise specified, the first of the two is the one being used.

The coinductive definition of bisimulation is equal to the standard one where bisimulation is regarded as the union of all bisimulation relations.
\begin{defi}A relation $\mathcal{R}$ is a bisumlation relation if for all $(P,~Q)\in\mathcal{R}$, $P\leadsto_\mathcal{R} Q$ and $Q\leadsto_\mathcal{R} P$. We define $\sim'$ to be the union of all bisumlation relations.
\end{defi}
\noindent We find the coinductive approach easier to work with and the proof that the two versions of bisimilarity are equal is straightforward.
\begin{lem}$\mathrel{\sim}\; \mathrel{=}\; \mathrel{\sim'}$\end{lem}
\proof\hfil
\begin{enumerate}[$\Rightarrow$]
\item By definition of $\sim$ we get for all processes $P$ and $Q$ where $P \sim Q$ that $P \leadsto_\sim Q$ and $Q\leadsto_\sim P$. Hence $\sim$ is a bisumlation relation.
\item[$\Leftarrow$] From the definition of $\sim'$ we get an arbitrary bisimulation relation $\mathcal{R}$ and processes $P$ and $Q$ where $(P,~Q)\in\mathcal{R}$, $P\leadsto_\mathcal{R}Q$ and $Q\leadsto_\mathcal{R}$. That $P\sim Q$ follows immediately by coinduction using lemma \ref{bisimWeakCoinduct} where $\mathcal{X}$ is set to $\mathcal{R}$.\qed
\end{enumerate}

\noindent An important property of the bisimulation relation is that it is equivariant. When doing proofs we rely heavily on Lemma \ref{simintro} which requires the simulation relation to be equivariant.

\begin{lem}
\label{bisimEqvt}
$\mbox{\rm eqvt}~\sim$
\end{lem}

\proof
By coinduction using Prop. \ref{bisimCoinduct} on $\sim$. Set $\mathcal{X}$ to be $\{(p\bullet P,~p\bullet Q)~|~P\sim Q\}$. Using Lemma \ref{simEqvt} the proof is quite straight forward since $\sim$ is a subset of $\mathcal{X}$ by instantiating $\mathcal{X}$ with the identity permutation. $\mathcal{X}$ is also trivially equivariant.\qed

  Another important property of strong bisimulation is that it is an equivalence relation.

\begin{lem}$\sim$ is an equivalence relation.\end{lem}
\label{bisimEquivalence}
\proof\hfil
\begin{enumerate}[\hbox to9 pt{\hfill}]
\item\noindent{\hskip-14 pt\emph{Reflexivity}:}\ Use coinduction and set $\mathcal{X}$ to the identity relation. The proof then follows trivially from Lemma \ref{simRefl}.
\item\noindent{\hskip-14 pt\emph{Symmetry}:}\ Follows trivially from the definition of $\sim$.
\item\noindent{\hskip-14 pt\emph{Transitivity}:}\ By coinduction where $\mathcal{X}$ is set to $\sim\circ\sim$. The result then follows by using Lemma \ref{simTrans}.\qed
\end{enumerate}
\pagebreak
\noindent We can now prove one of our main theorems.
\begin{thm}Strong bisimulation is preserved by all operators except the input-prefix, i.e.\\
\begin{center}
$
\begin{array}{@{}r@{\; \; \; }rlll}
\mathit{if}\; P\sim Q\; \mathit{then}&\tau.P&\sim&\tau.Q&(1)\\
\mathit{and}&\bar{a}b.P&\sim&\bar{a}b.Q&(2)\\
\mathit{and} &[a=b]P&\sim&[a=b]Q&(3)\\
\mathit{and} &[a\neq b]P&\sim&[a\neq b]Q&(4)\\
\mathit{and} &P + R&\sim&Q + R&(5)\\
 \mathit{and}&P\parop R&\sim&Q\parop R&(6)\\
\mathit{and} &(\nu x)P&\sim&(\nu x)Q&(7)\\
\mathit{and} &!P&\sim&!Q&(8)\\
\end{array}
$
\end{center}
\label{bisimPres}
\end{thm}
\begin{proof}
To prove (1) to (5), Lemmas \ref{tauPres}-\ref{outputPres} and \ref{matchPres}-\ref{sumPres} are used respectively.

When proving (7) we use coinduction and set $\mathcal{X}$ to $\{((\nu x)P,~(\nu x)Q)~|~P\sim Q\}$. Lemma \ref{resPres} can then prove preservation of both simulations. 

To prove (8) we strengthen our assumption that $P\sim Q$ to $(P,~Q)\in\mbox{\rm Rep}~\sim$ and use the coinduction principle \ref{bisimWeakCoinduct} with $\mathcal{X}$ set to $\mbox{\rm Rep}~\sim$. The preservation properties of the simulations can then be inferred by induction over $\mbox{\rm Rep}~\sim$ resulting in three cases from the derivation rules of $\mbox{\rm Rep}$. These can be proven by Lemmas \ref{bangPres}, \ref{parPres} and \ref{resPres} respectively.

The proof for (6) is deferred to Chapter \ref{exampleDeriv}.
\end{proof}

\noindent We now define strong equivalence as the largest bisimulation relation closed under substitution and prove our next theorem.
\begin{thm}
\label{eqCong}
$\sim^s$ is a congruence.
\end{thm}
\noindent This result uses Theorem \ref{bisimPres}. In the preservation proof for the $\nu$-operator the bound name must be $\alpha$-converted to not clash with the substitution chain. We also need the following lemma to prove closure under input-prefix.
\begin{lem}
$P\sim^s Q\Longrightarrow a(x).P\sim^s a(x).Q$
\end{lem}
\begin{proof}
By the definition of $\sim^s$ and Lemma \ref{inputPres}.
\end{proof}

\section{An Example Derivation}
\label{exampleDeriv}
\noindent As an example of our proof techniques, we here present the omitted part of the proof for Theorem \ref{bisimPres}(6) -- that strong bisimulation is preserved by the parallel operator.

The proof strategy amounts to proving simulations
$P\leadsto_\mathcal{R}Q$. We begin by stating the
requirements on $\mathcal{R}$ that are necessary for the proof to go
through. We do this before instantiating $\mathcal{R}$, since
this makes the proof more general and better structured.

Recall Lemma \ref{parPres} which is our preservation result for the $|$-operator.
\[
\begin{array}{rc}
(1)&  P \leadsto_\mathcal{R} Q\\
(2)&(P,~Q)\in\mathcal{R}\\
(3)&\mbox{\rm Id}\subseteq\mathcal{R}\\
(4)&  \forall P~Q~S.~(P,~Q) \in \mathcal{R}\Longrightarrow(P~|~S,~Q~|~S)\in\mathcal{R'}\\
(5)&  \forall P~Q~x.~(P,~Q) \in \mathcal{R'}\Longrightarrow((\nu x)P,~(\nu x)Q)\in\mathcal{R'}\\
\hline
\nonumber ~&P~|~S \leadsto_{\mathcal{R'}} Q~|~S
\end{array}
\]
Two of these conditions concern $\mathcal{R'}$. Condition (4) is straightforward -- if $P$
and $Q$ are in $\mathcal{R}$, then $P~|~S$ and $Q~|~S$ must be in $\mathcal{R'}$. Condition (5) is a bit less obvious but since the
parallel operator can introduce restrictions, $\mathcal{R'}$ must also be
preserved by the $\nu$-operator. Assumptions (2) and (3) ensure that the processes are in the $\mathcal{R}$ to begin with. This is not a prerequisite for simulation, but we need to know this in order to use (1) when a process stands still and we need to place it in parallel with the derivative of the other process in $\mathcal{R'}$.

We provide a more in depth look at the proof for Lemma \ref{parPres}.
\begin{proof}By Definition \ref{simulation} we shall show:\\
\[
\begin{array}{@{}ll}
(\forall a~x~T'.&Q~|~S\gta{a\ll x\gg}T' \land x~\sharp~P~|~S \Longrightarrow\\
~&\exists P'.~P~|~S\gta{a\ll x\gg}P' \land \mbox{\rm derivative}(a,~x,~P',~T',~\mathcal{R'}))~\land
\end{array}\\
\]
\[
(\forall \mathit{T'}~\alpha.~Q~|~S\gta{\alpha}\mathit{T'}\Longrightarrow\exists P'.~P~|~S\gta{\alpha}P'\land(P',~\mathit{T'})\in\mathcal{R'})
\]\\
We can now do case analysis on $Q~|~S\gta{\alpha}\mathit{T'}$ and $Q~|~S\gta{a\ll x\gg}T'$. We get eight cases (the four rules for parallel composition as seen in Fig.\ \ref{opSem} and their symmetric versions). We will focus on the \emph{Close}-case, as it nicely demonstrates the advantages of the nominal package. Using our derived case analysis rule, Lemma \ref{parStruct}, we can make sure that the bound names which appear in the \emph{Close}-case do not clash with $P$ by setting $\mathcal{C}$ to $P$. After induction we get:\\
\begin{center}
\begin{tabular}{lll}
(6)&$Q\gta{a(x)}Q'$&(assumption)\\
(7)&$S\gta{\bar{a}(y)}S'$&(assumption)\\
(8)&$x~\sharp~P$&($\mathcal{C}=P$ in Lemma \ref{parStruct})\\
(9)&$y~\sharp~P$&($\mathcal{C}=P$ in Lemma \ref{parStruct})\\
~\\
(10)&$\exists P'.~P\gta{a(x)}P'\land\mbox{\rm derivative}((\mbox{\rm Input}~a),~x,P',Q')$&(1, 6, 8, Def. \ref{simulation})\\
~\\
(11)&$P\gta{a(x)}P'$&(10)\\
(12)&$P~|~S\gta{\tau}(\nu y)(P'\{y/x\},~S')$&(\emph{Close}, 11, 7, 9)\\
~\\
(13)&$(P'\{y/x\},~Q'\{y/x\})\in\mathcal{R}$&(10, Def. \ref{simulation})\\
(14)&$(P'\{y/x\}~|~S',~Q'\{y/x\}~|~S')\in\mathcal{R'}$&(4, 13)\\
(15)&$((\nu y)(P'\{y/x\}~|~S'),~(\nu y)(Q'\{y/x\}~|~S')\in\mathcal{R'})$&(5, 14)\\
\hline
~&$\exists P'.~P~|~S\gta{\tau}P' \land (P',~(\nu y)(Q'\{y/x\}~|~S'))\in\mathcal{R'}$&(12, 15)\\
\end{tabular}
\end{center}
\end{proof}
\noindent The above is a step-by-step version of the Isabelle proof and it
mimics the way one could do a strict pen-and-paper version of the proof.  Note how in
steps 6 and 7, the bound names of both transitions generated by the
induction rule are set to be fresh for $P$. We would otherwise have to
$\alpha$-convert both transitions. As it
stands, all $\alpha$-conversions are abstracted away completely. Steps
13-15 uses the preservation properties of $\mathcal{R}$ and
$\mathcal{R'}$ to prove that the proper derivatives are in
$\mathcal{R'}$.

Furthermore, we have to prove a lemma on chains of
restrictions, since the \emph{Close}-operator introduces
new restrictions, as was also seen in lemma \ref{parCasesF}.

\begin{defi}
$(\nu\tilde{v})P$ denotes a chain of restrictions applied to $P$ where $\tilde{v}$ is a list, possibly empty, of restrictions.
\end{defi}

\begin{lem}Introduction rule for restriction chains:\\
\label{resChain}
\[
\rulee{
P \leadsto_\mathcal{R} Q\\ \mbox{\rm eqvt}~\mathcal{R}\\
\forall P~Q~x.\ (P,~Q)\in\mathcal{R}\Longrightarrow((\nu x)P,~(\nu x)Q)\in\mathcal{R}
}{(\nu\tilde{v})P \leadsto_\mathcal{R} (\nu\tilde{v})Q}{}
\]
\end{lem}
\begin{proof}By induction on $\tilde{v}$.\end{proof}
\noindent The intuition behind the lemma is quite simple. If a simulation
relation $\mathcal{R}$ is preserved by the $\nu$-operator and $P$ simulates $Q$ preserving $\mathcal{R}$, then since $\mathcal{R}$ is preserved by restriction and thus $(\nu x)P$ simulates $(\nu x)Q$ preserving $\mathcal{R}$ for an arbitrary name $x$, then by induction $(\nu\tilde{v})P$ must simulate $(\nu\tilde{v})Q$ preserving $\mathcal{R}$ where $\tilde{v}$ is
an arbitrary chain of restricted names. This is a general lemma which
is used repeatedly when proving bisimulations using the parallel operator.

We now proceed to the main proof of Theorem \ref{bisimPres}(6) using coinduction. We will need a set $\mathcal{X}$ which captures the agents we are interested in and prove the simulations which compose the bisimulation. We define $\mathcal{X}$ as $\{((\nu \tilde{v})(P~|~R),~(\nu\tilde{v})(Q~|~R))~|~P\sim Q\}$. The two simulation proofs we use reside in our main lemma since they share the same assumption, which is the way the proof is done inside Isabelle.

\begin{proof}If $P\bisim Q$ then $\para{P}{R}\bisim\para{Q}{R}$.\\
\begin{tabular}{lll}
(1)&$P\sim Q$&(assumption)\\
(2)&$(P~|~R,~Q~|~R)\in\mathcal{X}$&(1, def. of $\mathcal{X}$)\\
\end{tabular}\\

In order to use coinduction using Prop.\ \ref{bisimCoinduct} we must prove that every pair in
$\mathcal{X}$ simulates preserving $\mathrel{\mathcal{X}}\cup\mathrel{\sim}$. The members
of $\mathcal{X}$ have chains of restrictions so we first have to use
Lemma \ref{parCasesF} with a specific simulation relation in order to
reason about them.
\begin{lem}
\label{bisimParc1}
if $P \leadsto_\sim Q$ then $P~|~R\leadsto_{\mathcal{X}\cup\sim}Q~|~R$
\end{lem}
\begin{proof}
\[\begin{tabular}{rll}
(i)&$P \leadsto_\sim Q$&(assumption)\\
(ii)&
$
\forall P~Q~R.~(P, Q)\in\; \sim~\Longrightarrow(P~|~R,~Q~|~R)\in\mathcal{X}\; \cup\sim
$& (Def. of $\mathcal{X}$)\\
(iii)&$\forall P~Q~x.~(P,~Q)\in\mathcal{X}\; \cup\sim~\Longrightarrow$
&(Def. of $\mathcal{X}$, Lemma \ref{bisimPres})\\
~&$((\nu x)P,~(\nu x)Q)\in\mathcal{X}\; \cup\sim$\\
\hline
~&$P~|~R\leadsto_{\mathcal{X}\cup\sim}Q~|~R$&(Lemma \ref{parCasesF}, i-iii, 1)
\end{tabular}
\]
\end{proof}
\noindent From this lemma we see why Lemma \ref{parPres} has to have different relations in the assumptions and the conclusion. The simulation we can assume is $P\leadsto_\sim Q$ but the one we need to prove is $P~|~R\leadsto_{\mathcal{X}\cup\sim}Q~|~R$.

We can now extend our simulation to include chains of restrictions.
\begin{lem}
\label{bisimParc2}
If $P \leadsto_\sim Q$ then $(\nu \tilde{v})(P~|~R)\leadsto_{\mathcal{X}\cup\sim}(\nu\tilde{v})(Q~|~R)$
\end{lem}
\proof
\[\begin{tabular}{rll}
(i)&$P\leadsto_\sim Q$&(assumption)\\
(ii)&$P~|~R\leadsto_{\mathcal{X}\cup\sim}Q~|~R$&(Lemma \ref{bisimParc1}, i)\\
(iii)&$\mbox{\rm eqvt}(\mathcal{X}\; \cup\sim)$&(def. of $\mathcal{X}$, Lemma \ref{bisimEqvt})\\
(iv)&$\forall P~Q~x.~(P, Q)\in\mathcal{X}\; \cup\sim\Longrightarrow$
&(def. of $\mathcal{X}$, Lemma \ref{bisimPres})\\
~&$((\nu x)P,~(\nu x)Q)\in\mathcal{X}\; \cup\sim$\\
\hline
~&$(\nu \tilde{v})(P~|~R)\leadsto_{\mathcal{X}\cup\sim}(\nu\tilde{v})(Q~|~R)$&(Lemma \ref{resChain}, ii-iv)
\end{tabular}
\]
\end{proof}
\noindent We can now prove our goal:
$$P~|~R\sim Q~|~R\qquad\hbox{(coinduction, 2, Lemma \ref{bisimParc2},
  Def. \ref{bisimulation})}\eqno{\qEd}
$$

\noindent It is interesting to note that we only have to prove simulations one
way. When set up this way, Isabelle manages the symmetric versions of
the proofs automatically. Of course, if the relation is not symmetric, such as in the proof of $(\nu x)P\sim P$
if $x~\sharp~P$, the two different directions require separate proofs, just as  when doing the proofs on paper.

\section{Structural congruence}
\label{kapStructCong}
\noindent Structural congruence rules are used to equate processes which are structurally different but intuitively behave in the same way. The way these rules are implemented differ in different formalisations. A common approach is to let the labeled transition system replace a term for a structurally congruent one in order to enable transitions. Another approach, and the one that we have chosen, is to prove that all structurally congruent terms are also bisimilar. The rules for structural congruence can be found in Fig.\ \ref{structCong}.

\begin{thm}
\label{congBisim}
If $P\equiv Q$ then $P\sim Q$.
\end{thm}
\noindent As in the previous section we need to create auxiliary lemmas for all simulations we are interested in. Proving Theorem \ref{congBisim} requires that every structural congruence rule is proven individually. We will here demonstrate the most complicated example which is to prove associativity of the $|$-operator. We will need the following two lemmas for simulation.
\begin{lem}
\label{parAssocLeft}
\[
\infer{(P\parop Q)\parop R\leadsto_\mathcal{R}P\parop(Q\parop R)}{
\begin{array}{l}
\forall P~Q~R.~((P\parop Q)\parop R,~P\parop(Q\parop R))\in\mathcal{R}\\
\forall P~Q~x.~(P, Q)\in\mathcal{R}\Longrightarrow((\nu x)P,~(\nu x)Q)\in\mathcal{R}\\
\forall P~Q~R~x.~x\fresh P\Longrightarrow((\nu x)((P\parop Q)\parop R),~P\parop (\nu x)(Q\parop R))\in\mathcal{R}\\
\forall P~Q~R~x.~x\fresh R\Longrightarrow(((\nu x)(P\parop Q))\parop R,~(\nu x)(P\parop(Q\parop R)))\in\mathcal{R}
\end{array}
}
\]
\end{lem}
\begin{proof}By case analysis over the $|$-operator. This proof contains 18 cases. The proofs individually are not very hard, there are just a lot of cases to cover. The assumptions used about the relation $\mathcal{R}$ are used extensively in the proof.
\end{proof}

\begin{lem}
\label{parAssocRight}
\[
\infer{P\parop(Q\parop R)\leadsto_\mathcal{R}(P\parop Q)\parop R}{
\begin{array}{l}
\forall P~Q~R.~(P\parop (Q\parop R),~(P\parop Q)\parop R)\in\mathcal{R}\\
\forall P~Q~x.~(P, Q)\in\mathcal{R}\Longrightarrow((\nu x)P,~(\nu x)Q)\in\mathcal{R}\\
\forall P~Q~R~x.~x\fresh P\Longrightarrow(P\parop (\nu x)(Q\parop R), (\nu x)((P\parop Q)\parop R))\in\mathcal{R}\\
\forall P~Q~R~x.~x\fresh R\Longrightarrow((\nu x)(P\parop(Q\parop R)),~((\nu x)(P\parop Q))\parop R)\in\mathcal{R}\\
\end{array}
}
\]
\end{lem}

\begin{proof}
Similar to Lemma \ref{parAssocLeft}.
\end{proof}

\begin{figure}
\fbox{\begin{minipage}{\hsize}
The structural congruence $\equiv$ is defined as the smallest congruence
satisfying the following laws:
\begin{enumerate}
\item If $P$ and $Q$ are variants of $\alpha$-conversion then $P \equiv Q$.
\item The abelian monoid laws for Parallel:
commutativity $P\parop Q \equiv Q\parop P$, associativity $(P\parop Q)\parop R \equiv P\parop (Q\parop R)$, and
$\nil$ as unit $P\parop \nil \equiv P$; and the same laws for Sum.
\item The unfolding law $!P\equiv P\parop!P$
\item The scope extension laws
\[\begin{array}{lcll}
(\nu x) \nil & \equiv & \nil\\
(\nu x)(P\;\parop \;Q)    &\equiv& P\;\parop \;(\nu x) Q  & \mbox{if $x\fresh P$}\\
(\nu x)(P+Q)&\equiv&P+(\nu x)Q  & \mbox{if $x\fresh P$}\\
(\nu x)[u=v]P& \equiv &[u=v](\nu x)P &
                              \mbox{if $x \neq u$ and $x \neq v$} \\
(\nu x)[u\neq v]P &\equiv& [u\neq v](\nu x)P &
                              \mbox{if $x\neq u$ and $x \neq v$} \\
(\nu x)(\nu y)P &\equiv& (\nu y)(\nu x)P
\end{array}
\]
\end{enumerate}
\caption{The definition of structural congruence.}
\label{structCong}
\end{minipage}}
\end{figure}
\noindent In order to do the rest of this proof efficiently it turns out that we need to use other rules for structural congruence since Lemma \ref{parAssocLeft} and \ref{parAssocRight} make heavy use of scoping rules. The coinduction rule (Prop. \ref{bisimCoinduct}) allows us to work with an arbitrary relation, but to include the laws of structural congruence in this relation would be cumbersome. Instead we create the following coinduction rule.
\begin{lem}Compositional coinduction rule. Let $\mathcal{Y}$ be $\sim\circ\; (\mathcal{X}\; \cup\sim)\; \circ\sim$.
\label{compBisimCoinduct}
\[
\rulee{
(P,~Q)\in\mathcal{X} \and
\mbox{\rm eqvt}~\mathcal{X} \\\\
\forall P'~Q'.~(P',~Q')\in~\mathcal{X}\Longrightarrow P'\leadsto_{\mathcal{Y}} Q'\wedge Q'\leadsto_\mathcal{Y} P'
}
{P\sim Q}{}
\]
\end{lem}
\begin{proof}By coinduction and transitivity of simulation.\end{proof}
We can now prove associativity of the $|$-operator.
\begin{lem}$(P\parop Q)\parop R\; \mathrel{\sim}P\parop(Q\parop R)$\end{lem}
\begin{proof}
By coinduction using Lemma \ref{compBisimCoinduct} and setting $\mathcal{X}$ to\\ $\{((\nu\tilde{v})((P\parop Q)\parop R),~(\nu\tilde{v})(P\parop(Q\parop R)))\}$. Lemma \ref{parAssocRight} and \ref{parAssocLeft} can then be used together with the laws for scope extrusion to complete the proof.
\end{proof}
The next step is to prove that all structurally congruent terms are strongly equivalent.
\begin{thm}
\label{congEq}
$P\equiv Q\Longrightarrow P\sim^s Q$
\end{thm}
\begin{proof}Nearly all work has already been done in Theorem \ref{congBisim}. These proofs do, however, require manual alpha conversions when dealing with scoping rules as the cases where the restricted name clashes with the substitution chain must be taken into consideration. This is an example of where pen-and-paper proofs often are less rigorous than strictly required.

\end{proof}

\noindent The proofs we have done in this section are not overly complicated but require a solid attention to detail. Many of the proofs have many cases and even though the results have never been in doubt, having them fully machine checked convinces us that no case has been overlooked. Moreover, without the framework to abstract away from bound names the amount of cases for all different $\alpha$-variants would have been very much larger. 

\section{Weak bisimulation}
\label{kapWeakBisim}
\subsection{Basic definitions}
Weak bisimulation equivalence is often called observation equivalence. The intuition is that $\tau$-transitions are considered internal and hence invisible to the outside environment. For two processes to be observation equivalent, they only need to mimic the visible actions of each other. More formally, we reason about a $\tau$-chain $P\Gta{\widehat{\tau}}P'$ as the reflexive transitive closure of $\tau$-actions, i.e.
$P\Gta{\widehat{\tau}}P'\define P\gta{\tau}^*P'$. A weak transition is then said to be an action preceded and succeeded by a $\tau$-chain.

Weak late bisimulation is complicated for input actions. It requires substitutions made as a result of the input to be
applied immediately to the input derivative before the succeeding
$\tau$-chain is executed, and that one such derivative can continue to simulate for all possible received names, see e.g. \cite{parrow:intro}. Therefore the weak late semantics needs to carry additional
information in the labels as follows.
\begin{defi}
\label{weakTransition}
\[
\begin{array}[t]{lll}
P\Gta{\alpha}P'&\define&P\Gta{\widehat{\tau}}\gta{\alpha}\Gta{\widehat{\tau}}P'\\
P\Gta{\bar{a}(x)}P'&\define&P\Gta{\widehat{\tau}}\gta{\bar{a}(x)}\Gta{\widehat{\tau}}P'\\
P \wli{u}{a(x)}{P''} P'&\define&P\Gta{\widehat{\tau}}\gta{a(x)}P''\wedge P''\{u/x\}\Gta{\widehat{\tau}}P'\\
\end{array}
\]
\noindent Residuals are written in the same way for weak as for strong transitions, except for the input case which is written $u:a(x)@P''\prec P'$. A transition can also be written as $P\Longmapsto\mbox{\rm Res}$ where $\mbox{\rm Res}$ is a residual.
\end{defi}
\noindent The transition $P\wli{u}{a(x)}{P''}P'$ means that $P$ can do a $\tau$-chain and then $a(x)$ to an agent $P''$ where $x$ is substituted for $u$ and another $\tau$-chain is done to $P'$. The agent $P''$ represents the exact state where the substitution is made. This will be important when we define weak simulation.

Note that the bound name $x$ in the bound output case is bound in $P'$ and normal $\alpha$-conversions can be applied. Also, even though we are modeling a late semantics, the name $x$ is \emph{not} bound in $P'$ in the input-transition as it is substituted for $u$ before the $\tau$-chain. We can still do $\alpha$-conversions through the following lemma:
\begin{lem}
if $P\wli{u}{a(x)}{P''}P'$ and $y~\sharp~P$ then $P\wli{u}{a(y)}{(x ~ y) \bullet P''}P'$
\end{lem}
\noindent We also need to weaken the transitions in the standard way:
\begin{defi}Weak late transitions
\label{weakTauTransition}
\[
\begin{array}[t]{lll}
P\Gta{\widehat{\alpha}}P'&\define&P\Gta{\widehat{\tau}}P'~\mbox{\rm if}~\alpha=\tau\\
~&~&P\Gta{\alpha}P'~\mbox{\rm otherwise}
\end{array}
\]
\end{defi}
\noindent We can now define weak late simulation.
\begin{defi}
\label{weakSimulation}
The agent $P$ can weakly late simulate the agent $Q$ preserving $\mathcal{R}$, written $P\weakSim{\mathcal{R}}Q$, if
\[
\begin{array}{@{}ll}
(\forall a~x~Q'.&Q\gta{\bar{a}(x)}Q' \wedge x~\sharp~P \Longrightarrow\\
~&\exists P'.~P\Gta{\bar{a}(x)}P' \wedge (P',~Q') \in \mathcal{R})~\wedge
\end{array}
\]
\[
\begin{array}{@{}ll}
(\forall a~x~Q'.&Q\gta{a(x)}Q' \wedge x~\sharp~P \Longrightarrow\\
~&\exists P''.~\forall u.~\exists P'.~P\wli{u}{a(x)}{P''}P' \wedge (P',~Q'\{u/x\}) \in \mathcal{R})~\wedge
\end{array}
\]
\[
(\forall \alpha~Q'.~Q\gta{\alpha}Q' \Longrightarrow \exists P'.~P\Gta{\widehat{\alpha}}P' \wedge (P',~Q') \in \mathcal{R})
\]
\end{defi}
\noindent The important aspect of weak late simulation is the fact mentioned above -- that an input-action $a(x)$ must be matched by a weak transition with the same input derivative $P''$ for \emph{all} possible instantiations $u$ of the bound name. From our definition, we can derive an introduction rule for weak simulation similar to the one done for strong simulation in Lemma \ref{simintro}.

In the standard way we define another version of simulation $\weakCongSim$ where we require the simulating process to do at least one action to mimic the simulated agent. The definition of $\weakCongSim$ is the same as for $\weakSim{}$ except that the simulating process in the last conjunct uses $\Gta{\alpha}$ instead of $\Gta{\widehat{\alpha}}$.
\begin{defi}
\label{weakCongSimulation}$P\weakCongSim_\mathcal{R} Q$ if\\
\[
\begin{array}{@{}ll}
(\forall a~x~Q'.&Q\gta{\bar{a}(x)}Q' \wedge x~\sharp~P \Longrightarrow\\
~&\exists P'.~P\Gta{\bar{a}(x)}P' \wedge (P',~Q') \in \mathcal{R})~\wedge
\end{array}
\]
\[
\begin{array}{@{}ll}
(\forall a~x~Q'.&Q\gta{a(x)}Q' \wedge x~\sharp~P \Longrightarrow\\
~&\exists P''.~\forall u.~\exists P'.~P\wli{u}{a(x)}{P''}P' \wedge (P',~Q'\{u/x\}) \in \mathcal{R})~\wedge
\end{array}
\]
\[
(\forall \alpha~Q'.~Q\gta{\alpha}Q' \Longrightarrow \exists P'.~P\Gta{\alpha}P' \wedge (P',~Q') \in \mathcal{R})
\]
\end{defi}\subsection{Lifted semantics}
Our preservation proofs for weak transitions are very similar to the corresponding proofs for strong transitions. We achieve this by \emph{lifting} the operational semantics, i.e. mapping each rule from Fig. \ref{opSem} to a corresponding rule using weak transitions. The following transition system can be derived for the transitions defined in Def. \ref{weakTransition}.
\begin{lem}\label{weakOpsem} The lifted semantics for the transitions 
defined in Def. \ref{weakTransition}.
\begin{mathpar}
\rulee{}{a(x).P\wli{u}{a(x)}{P}P\{u/x\}}{Input~~~}\and
\rulee{}{\bar{a}b.P\Gta{\bar{a}b}P}{Output~~~}\and
\rulee{}{\tau.P\Gta{\tau}P}{Tau~~~}\and
\rulee{P\Longmapsto\mbox{Res}}{[a=a]P\Longmapsto\mbox{Res}}{Match}\and
\rulee{P\Longmapsto\mbox{Res}~~~a\neq b}{[a\neq b]\Longmapsto\mbox{Res}}{Mismatch}\and
\rulee{P\Gta{\bar{a}b}P'~~~a\neq b}{(\nu b)P\Gta{\bar{a}(b)}P'}{Open}\and
\rulee{P\Longmapsto\mbox{Res}}{P + Q\Longmapsto\mbox{Res}}{Sum}\and
\rulee{P\wli{u}{a(x)}{P''}P'~~~~x~\sharp~Q}{P~|~Q \wli{u}{a(x)}{P''|Q} P'~|~Q}{ParIn}\and
\rulee{P\Gta{\bar{a}(x)}P'~~~~x~\sharp~Q}{P~|~Q\Gta{\bar{a}(x)}P'~|~Q}{ParBO}\and
\end{mathpar}
\begin{mathpar}
\rulee{P\Gta{\alpha}P'}{P~|~Q\Gta{\alpha}P'~|~Q}{ParF}\and
\rulee{P\wli{b}{a(x)}{P''}P'~~~Q\Gta{\bar{a}b}Q'}{P~|~Q\Gta{\tau}P'~|~Q'}{Comm}\and
\rulee{P\wli{y}{a(x)}{P''}P'~~~Q\Gta{\bar{a}(y)}Q'~~~y~\sharp~P}{P~|~Q\Gta{\tau}(\nu y)(P'~|~Q')}{Close}\and
\rulee{P\Gta{\bar{a}(x)}P'~~~y\fresh(a,~x)}{(\nu y)P\Gta{\bar{a}(x)}(\nu y)P'}{ResBO}\and
\rulee{P\wli{u}{a(x)}{P''}P'~~~y\fresh(a,~u,~x)}{(\nu y)P\wli{u}{a(x)}{(\nu y)P''}(\nu y)P'}{ResIn}\and
\rulee{P\Gta{\alpha}P'~~~x\fresh\alpha}{(\nu x)P\Gta{\alpha}(\nu x)P'}{ResF}\and
\rulee{P\parop!P\Longmapsto\mbox{Res}}{!P\Longmapsto\mbox{Res}}{Replication}
\end{mathpar}
\end{lem}

\noindent When trying to lift the semantics to the transitions defined in Def.\ \ref{weakTauTransition} we encounter difficulties. The rules which do not have $\Gta{\hat\alpha}$ in the assumptions trivially follow from Lemma \ref{weakOpsem}, but of the remaining, only $\mathbf{ParF}$ and $\mathbf{ResF}$ can be lifted.
\begin{cor}The lifted rules for $\mathbf{ParF}$ and $\mathbf{ResF}$.
\small
\label{weakTauOpsem}
\[
\rulee{P\Gta{\widehat{\alpha}}P'}{P~|~Q\Gta{\widehat{\alpha}}P'~|~Q}{ParF}
\qquad\qquad
\rulee{P\Gta{\widehat{\alpha}}P'~~~x\fresh\alpha}{(\nu x)P\Gta{\widehat{\alpha}}(\nu x)P'}{ResF}
\]
\end{cor}

\noindent The operational rules from Fig.\ \ref{opSem} that we cannot
be lifted in this manner, as opposed to the ones in Lemma
\ref{weakOpsem}, are $\mathbf{Match}$, $\mathbf{Mismatch}$,
$\mathbf{Sum}$ and $\mathbf{Replication}$ in the case where
$\alpha=\tau$ and $P=P'$.

\subsection{Preservation properties}
To prove preservation properties for weak simulations we need to lift the preservation proofs from strong simulations to weak ones. For $\weakCongSim$ this turns out to be unproblematic. The lemmas require the same assumptions to be proven with the addition that we sometimes need to know that if $(P,~Q)\in\mathcal{R}$ then $P\weakCongSim_\mathcal{R}Q$. The reason for this is that after following a $\tau$-chain, we need to know that we are still inside the simulation. For $\weakSim{}$, however, the lemmas that we could not lift in Cor.\ \ref{weakTauOpsem} need their assumption strengthened. These lemmas are:
\begin{lem}
\label{weakMatchPres}
\[
\rulee{P\weakSim{\mathcal{R}}Q\\\\
\forall P~Q~a.~(P,~Q)\in\mathcal{R}\Longrightarrow([a=a]P,~Q)\in\mathcal{R}
}
{[a=b]P\weakSim{\mathcal{R}}[a=b]Q}
{}
\]
\end{lem}
\begin{proof}
By the definition of $\weakSim{}$ and Prop.\ \ref{matchCases}. In the case where the $\tau$-transition stands still, the second assumption is used to prove that the derivatives are still in $\mathcal{R}$.
\end{proof}
\begin{lem}
\label{weakMismatchPres}
\[
\rulee{
P\weakSim{\mathcal{R}}Q\\\\
\forall P~Q~a~b.~(P,~Q)\in\mathcal{R}\wedge a\neq b\Longrightarrow([a\neq b]P,~Q)\in\mathcal{R}
}
{[a\neq b]P\weakSim{\mathcal{R}}[a\neq b]Q}
{}
\]
\end{lem}
\begin{proof}
By the definition of $\weakSim{}$ and Prop.\ \ref{mismatchCases}. In the case where the $\tau$-transition stands still, the second assumption is used to prove that the derivatives are still in $\mathcal{R}$.
\end{proof}
\begin{lem}
\label{weakBangPres}
\[
\rulee{
(P,~Q)\in\mathcal{R}\and
\mbox{\rm eqvt}~\mathcal{R}\\\\
\forall P~Q.~(P,~Q)\in\mathcal{R}\Longrightarrow P\leadsto_\mathcal{R} Q\\\\
\forall P~Q.~(P\parop!P,~Q)\in\mathcal{R}\Longrightarrow (!P, Q)\in\mathcal{R}
}
{!P\leadsto_{\mbox{\rm Rep}~\mathcal{R}} !Q}
{}
\]
\end{lem}
\begin{proof}
Similar to Lemma \ref{bangPres} but when a $\tau$-action stands still the fourth assumption is used.
\end{proof}
\noindent The other preservation lemmas look the same as their strong counterparts. Their proofs need to treat input-actions differently as there is a noticeable difference in how input-actions are treated in strong and weak simulations. Other than this, the proofs follow the same pattern.

Weak bisimulation equivalence is defined using coinduction in exactly the same way as strong bisimulation. As a result, all coinduction rules which were generated for strong bisimulation are also generated for weak.

\begin{defi}\label{weakBisimulation} Weak bisimulation equivalence, 
$\approx$, is the largest relation satisfying:
\[P\weakBisim Q \Longrightarrow
P\weakSim{\weakBisim}Q\mathrel{\wedge}Q\weakSim{\weakBisim}P
\]
\end{defi}
\noindent Weak bisimulation is not a congruence since it is neither preserved by the $+$-operator nor by the input-prefix, but it is preserved by all other operators.
\begin{thm}
\label{weakBisimPres}
$\approx$ is preserved by all operators except $+$ and input prefix.
\end{thm}
\begin{proof}
The first step in in this proof is to use the lifted preservation rules for weak simulation. In order to prove preservation of $\mathbf{Match}$, $\mathbf{Mismatch}$ and $\mathbf{Replication}$, we need the results $P\approx[a=a]P$, $P\approx[a\neq b]P$ when $a\neq b$ as well as the structural congruence result $P~|~!P\approx~!P$.
\end{proof}

\noindent To obtain a congruence we follow the standard procedure.
The proofs of the preservation lemmas for $\weakCongSim$ are similar to their strong counterparts since all rules from the operational semantics can be lifted using Lemma \ref{weakOpsem}.

We can now define weak congruence.
\begin{defi}
\label{weakEquivalence}
$P \weakCong Q \define P~\weakCongSim_\approx~Q~\wedge~Q~\weakCongSim_\approx~P$
\end{defi}
\noindent Note that this is not a coinductive definition since it refers to
$\approx$. The proof that $\weakCong$ is preserved by all operators except input-prefix corresponds closely to our corresponding proof for $\sim$. The proof that $\cong^s$ is a congruence follows in the same manner.
\begin{thm}
\label{weakEqPres}
$\weakCong$ is preserved by all operators except input-prefix.
\end{thm}
\begin{proof}
This proof is nearly identical to the one for Theorem \ref{bisimPres}, but we use our preservation proofs for $\weakCongSim$ instead of the ones for $\leadsto$.
\end{proof}
\begin{lem}\label{weakEqCong}$\weakCong^s$ is a congruence\end{lem}
\begin{proof}
Similar to Lemma \ref{eqCong}.
\end{proof}
\subsection{Relationships between equivalences}
\label{weakLateStructCong}

We prove that $\bisim\; \subseteq\; \weakCong\; \subseteq\; \weakBisim$. Among other things, this implies that the weaker bisimlation equivalences contain structural congruence.
The first part of this proof is to establish  correspondance properties between the different types of transitions.
\begin{cor}
\label{transitionWeakStepTransition}
\[
\begin{array}[t]{l@{\; }l}
\mbox{If}\; P\gta{\bar{a}(x)}P'\; \mbox{then}\; P\Gta{\bar{a}(x)}P'\\
\mbox{If}\; P\gta{\alpha}P'\; \mbox{then}\; P\Gta{\alpha}P'\\
\mbox{If}\; P\gta{a(x)}P'\; \mbox{then}\; P\wli{u}{a(x)}{P'}P'\{u/x\}
\end{array}
\]
\end{cor}
\begin{proof}
Follows from the definition of $P\Longmapsto\mathit{Res}$ by adding empty $\tau$-chains before and after the transitions.
\end{proof}
\noindent The next step is to do the same for simulations.
\begin{cor}
\label{simWeakCongSim}
If $P\leadsto_\mathcal{R}Q$ then $P\weakCongSim_\mathcal{R} Q$
\end{cor}
\begin{proof}
By the definition of $\leadsto$, $\weakCongSimOp$\ and Cor.\ \ref{transitionWeakStepTransition}.
\end{proof}
\noindent And finally for weak congruence.
\begin{cor}
\label{bisimWeakCong}
If $P\bisim Q$ then $P\weakCong Q$
\end{cor}
\begin{proof}
By the definition of $\bisim$, $\weakCong$\ and Cor.\ \ref{simWeakCongSim}.
\end{proof}
\noindent The corresponding proof for our congruence relations follow trivially.
\begin{cor}
\label{bisimSubstWeakCongSubst}
If $P\bisim^s Q$ then $P\weakCong^s Q$
\end{cor}
\begin{proof}
 Follows from the definitions of $\bisim^s$, $\weakCong^s$ and Cor.\ \ref{bisimWeakCong}.
\end{proof}
\noindent We can use the same technique when reasoning about weak bisimulation.
\begin{cor}
\label{weakStepTransitionWeakTransition}
If $P\Gta{\alpha}P'$ then $P\Gta{\widehat{\alpha}} P'$
\end{cor}
\begin{proof}
Follows from the definitions of $\Gta{\alpha}$ and $\Gta{\widehat{\alpha}}$ as $\Gta{\alpha}$ can do everything $\Gta{\widehat{\alpha}}$ can do except doing an empty sequence of $\tau$s.
\end{proof}
\noindent Followed by simulation
\begin{cor}
\label{weakStepSimWeakSim}
If $P\weakCongSim Q$ then $P\weakSim{} Q$
\end{cor}
\begin{proof}
Follows from the definitions of $\weakCongSim$, $\weakSimOp$ and Cor.\ \ref{weakStepTransitionWeakTransition}.
\end{proof}
\noindent And finaly for weak bisimulation.
\begin{cor}
\label{weakCongWeakBisim}
If $P\cong Q$ then $P\approx Q$
\end{cor}
\begin{proof}
Follows from the definitions of $\approx$, $\weakCong$ and Cor.\ \ref{weakStepSimWeakSim}.
\end{proof}
\noindent Using the techniques above our results follow as a simple corollary.
\begin{cor}
\[
 \begin{array}[t]{l@{\; }r@{\; }l}
  \mbox{If}\; P\equiv Q&\mbox{then}&P\approx Q\\
                 &\mbox{and} &P\weakCong Q\\
                 &\mbox{and} &P\weakCong^s Q
 \end{array}
\]
\end{cor}
\begin{proof}
 Follows from Theorem \ref{congEq} and Corollaries \ref{bisimWeakCong}, \ref{weakCongWeakBisim} and \ref{bisimSubstWeakCongSubst}.
\end{proof}

\subsection{The Hennessy Lemma}
As an example we prove the Hennessy Lemma.
\begin{thm}$P\approx Q$ iff $\tau.P\cong Q \vee P \cong Q \vee P\cong\tau.Q$
\end{thm}
\begin{proof}
We first prove the lemma in the direction left-to-right. We will need the following auxiliary lemmas.
\begin{lem}
\label{hennessyRight1}
If $P\weakSim{\mathcal{R}}Q$ then $\tau.P\weakCongSim_\mathcal{R} Q$
\end{lem}
\begin{proof}
By the definition of $\weakCongSim$. The interesting case is the $Q$ does a $\tau$-action, and $\tau.P$ can always mimic that with at least one step since $P\weakSim{\mathcal{R}}Q$.
\end{proof}
\begin{lem}
\label{hennessyRight2}
\[
\rulee{
P\weakSim{\mathcal{R}} Q\\\\
\forall Q'.~Q\gta{\tau} Q'\Longrightarrow (P,~Q')\notin\mathcal{R}
}
{P\weakCongSim_\mathcal{R} Q}{}
\]
\end{lem}
\begin{proof}This follows from the definition of $\weakSim{}$ and $\weakCongSim$. The only difference being that for a $\tau$-transition, the simulating process by $\weakSim{}$ can do an empty sequence of $\tau$s. whereas in $\weakCongSim$ it cannot. In our assumptions we remove this option.
\end{proof}
\begin{lem}
\label{hennessyRight3}
If $P\gta{\tau}P'$ and $(P', Q)\in\mathcal{R}$ then $P\weakCongSim_\mathcal{R}\tau.Q$
\end{lem}
\begin{proof}
Follows from the definition of $\weakCongSim$.
\end{proof}
\noindent We can now complete our proof for the left-to-right direction of the Hennessy lemma by doing proofs on the following cases:
\begin{enumerate}[(1)]
\item $\exists P'. P\gta{\tau}P'\wedge P'\approx Q$
\item $\exists Q'. Q\gta{\tau}Q'\wedge P\approx Q'$
\end{enumerate}
In the case that $1$ or $2$ holds we use Lemmas \ref{hennessyRight1} and \ref{hennessyRight3} to prove the first and third disjunct. In the case that neither hold, Lemma \ref{hennessyRight2} can be used for both directions of the bisimulation. This concludes the proof in the left-to-right direction.

We will need the following lemmas for the direction right-to-left.
\begin{lem}
\label{hennessyLeft1}
If $\tau.P\weakCongSim_\mathcal{R} Q$ then $P\weakSim{\mathcal{R}}Q$
\end{lem}
\begin{proof}
By the definition of $\weakSim{}$. If $Q$ does a $\tau$-action and $\tau.P$ simulates by doing a single $\tau$-step, $P$ can stand still and end up in the same state. Otherwise, $P$ can always move to the same state as $\tau.P$ by doing one less $\tau$-step.
\end{proof}
\begin{lem}
\label{hennessyLeft2}
\[
\rulee{P\weakCongSim_\mathcal{R}\tau.Q\\\\
\forall P'~Q'.~(P',~Q')\in\mathcal{R}\Longrightarrow P'\weakSim{\mathcal{R}} Q'
}
{P\weakSim{\mathcal{R}} Q}{}
\]
\end{lem}
\begin{proof}
From the definition of $\weakCongSim$ we get a $\tau$-chain $P\Longrightarrow_\tau P'$ for some $P'$ where $(P',~Q)\in\mathcal{R}$. We also know that $P'\weakSim{\mathcal{R}}Q$. By the definition of $\weakSim{}$ we can add the chain $P\Longrightarrow_\tau P'$ to any simulation of $Q$.
\end{proof}
\noindent To finish the the proof we use Lemmas \ref{hennessyLeft1} and \ref{hennessyLeft2} for the first and third disjunct and Cor.\
 \ref{weakCongWeakBisim} for the second one.
\end{proof}

\section{Early semantics and bisimulation}
\label{kapEarly}
\subsection{Early semantics}
In the early semantics the input action carries the name received rather than a bound name, so we have that the process $a(x).P$ can receive all names $u$ doing an action $au$ and ending up in the derivative $P\{u/x\}$. The main difference to late semantics is that substitution is done at the input prefix rule, i.e. as early as possible, and not during communication.

\begin{figure}
\fbox{\begin{minipage}{\hsize}
\begin{mathpar}
\rulee{}{a(x).P\gta{au}_eP\{u/x\}}{Input}\and
\rulee{}{\bar{a}b.P\gta{\bar{a}b}_eP}{Output}\and
\rulee{}{\tau.P\gta{\tau}_eP}{Tau}\and
\rulee{P\longmapsto_e\mbox{Res}}{[a=a]P\longmapsto_e\mbox{Res}}{Match}\and
\rulee{P\longmapsto_e\mbox{Res}~~~a\neq b}{[a\neq b]P\longmapsto_e\mbox{Res}}{Mismatch}\and
\rulee{P\gta{\bar{a}b}_eP'~~~a\neq b}{(\nu b)P\gta{\bar{a}(b)}_eP'}{Open}\and
\rulee{P\longmapsto_e\mbox{Res}}{P~+~Q\longmapsto_e\mbox{Res}}{Sum}\and
\rulee{P\gta{\bar{a}(x)}_eP'~~~~x~\sharp~Q}{P~|~Q\gta{\bar{a}(x)}_eP'~|~Q}{ParB}\hspace{10mm}\and
\rulee{P\gta{\alpha}_eP'}{P~|~Q\gta{\alpha}_eP'~|~Q}{ParF}\and
\rulee{P\gta{ab}_eP'~~~Q\gta{\bar{a}b}_eQ'}{P~|~Q\gta{\tau}_eP'~|~Q'}{Comm}\and
\rulee{P\gta{ay}_eP'~~~Q\gta{\bar{a}(y)}_eQ'~~~y~\sharp~P}{P~|~Q\gta{\tau}_e(\nu y)(P'~|~Q')}{Close}\and
\rulee{P\gta{\bar{a}(x)}_eP'~~~~y~\sharp~(a,~x)}{(\nu y)P\gta{\bar{a}(x)}_e(\nu y)P'}{ResB}\and
\rulee{P\gta{\alpha}_eP'~~~~y~\sharp~\alpha}{(\nu y)P\gta{\alpha}_e(\nu y)P'}{ResF}\and
\rulee{P~|~!P\longmapsto_e\mbox{Res}}{!P\longmapsto_e\mbox{Res}}{Replication}
\end{mathpar}
\caption{The \emph{Par}- and the \emph{Res}-rule in the early operational semantics are still split, but the input action contains no bound names. Symmetric versions have been elided.}
\label{earlyOpSem}
\end{minipage}}
\end{figure}
The way we write actions differ somewhat from the late semantics. We write the early transitions in a similar way, but with a subscript \emph{e} to differentiate them from the late ones. Moreover, In the early semantics, a transition $\gta{\alpha}_e$ can include an input-transition as it does not contain a bound name. The intuition is that an action is denoted $\alpha$ if it contains no binders. As a result, our Isabelle definition for early residuals need to be changed.
\begin{defi}The early residual datatype.
\begin{verbatim}
datatype freeRes = InputR name name
                 | OutputR name name
                 | TauR

nominal_datatype residual = BoundOutputR name "«name» pi"
                          | FreeR freeRes pi

\end{verbatim}
\end{defi}
\subsection{Early bisimulation}
The definition of early simulation is similar to its late counterpart. The difference between the two is that no distinction has to be made for the input-action as the substitution takes place before any communication is made.
\begin{defi}
\label{earlySimulation}
The agent $P$ can early simulate the agent $Q$ preserving $\mathcal{R}$, written ${P\leadsto_{e\mathcal{R}}Q}$, if
\[
(\forall a~x~Q'.~Q\gta{\bar{a}(x)}_eQ' \land x~\sharp~P \Longrightarrow\exists P'.~P\gta{\bar{a}(x)}_eP' \land (P', Q')\in\mathcal{R})\land
\]
\[
(\forall \alpha~Q'.~Q\gta{\alpha}_eQ' \Longrightarrow \exists P'.~P\gta{\alpha}_eP' \land (P', Q') \in \mathcal{R})
\]
\end{defi}
\noindent Bisimulation is again defined using our standard coinduction technique.
\begin{defi}Early bisimulation equivalence, $\sim_e$,  is the largest relation satisfying:
\label{bisimulation}
\[
  P \sim_e Q \Longrightarrow P \leadsto_{e\sim_e} Q~\land~Q\leadsto_{e\sim_e} P
\]
\end{defi}
\noindent All the preservation proofs and congruence results for late bisimulation have also been done for early. This did require creating rules for case analysis on the early operational semantics in a similar way as was done for late. We have created the library of preservation lemmas similar to the one for late semantics. This work was pretty straightforward and the two libraries work in the same way except for how they treat input actions. Once this was done, the proofs for early bisimulation were nearly identical to their late counterparts and required very little extra work.
\begin{thm}
$\sim_e$ is preserved by all operators except input-prefix.
\end{thm}
\begin{thm}
$\sim_e^s$ is a congruence.
\end{thm}
\subsection{Weak early bisimulation}
We have also proven our results for weak early bisimulation. We use the same technique as we did for weak late bisimulation by lifting the early operational semantics to a weak counterpart. The weak early operational semantics can be written on a simpler form, however, as weak early simulation does not require any knowledge of the point that a substitution was made. A weak early transition is hence written $\Gta{\alpha}_e$ or $\Gta{\widehat\alpha}_e$, where $\alpha$ is an arbitrary transition.
\begin{lem}The lifted semantics for the weak early operational semantics.\\
\small
\label{weakEarlyOpsem}
\begin{mathpar}
\rulee{}{a(x).P\Gta{au}_eP\{u/x\}}{Input~~~}\and
\rulee{}{\bar{a}b.P\Gta{\bar{a}b}_eP}{Output~~~}\and
\rulee{}{\tau.P\Gta{\tau}_eP}{Tau~~~}\and
\rulee{P\Longmapsto\mkern-8mu_e\; \mbox{Res}}{[a=a]P\Longmapsto\mkern-8mu_e\; \mbox{Res}}{Match}\and
\rulee{P\Longmapsto\mkern-8mu_e\; \mbox{Res}~~~a\neq b}{[a\neq b]\Longmapsto\mkern-8mu_e\; \mbox{Res}}{Mismatch}\and
\rulee{P\Gta{\bar{a}b}_eP'~~~a\neq b}{(\nu b)P\Gta{\bar{a}(b)}_eP'}{Open}\and
\rulee{P\Longmapsto\mkern-8mu_e\; \mbox{Res}}{P + Q\Longmapsto\mkern-8mu_e\; \mbox{Res}}{Sum}\and
\rulee{P\Gta{\bar{a}(x)}_eP'~~~~x~\sharp~Q}{P~|~Q\Gta{\bar{a}(x)}_eP'~|~Q}{ParB}\and
\rulee{P\Gta{\alpha}_eP'}{P~|~Q\Gta{\alpha}_eP'~|~Q}{ParF}\and
\rulee{P\Gta{a(b)}_eP'~~~Q\Gta{\bar{a}b}_eQ'}{P~|~Q\Gta{\tau}_eP'~|~Q'}{Comm}\and
\rulee{P\Gta{a(y)}_eP'~~~Q\Gta{\bar{a}(y)}_eQ'~~~y~\sharp~P}{P~|~Q\Gta{\tau}_e(\nu y)(P'~|~Q')}{Close}\and
\rulee{P\Gta{\bar{a}(x)}_eP'~~~y\fresh(a,~x)}{(\nu y)P\Gta{\bar{a}(x)}_e(\nu y)P'}{ResB}\and
\rulee{P\Gta{\alpha}_eP'~~~x\fresh\alpha}{(\nu x)P\Gta{\alpha}_e(\nu x)P'}{ResF}\and
\rulee{P\parop!P\Longmapsto\mkern-8mu_e\; \mbox{Res}}{!P\Longmapsto\mkern-8mu_e\; \mbox{Res}}{Replication}
\end{mathpar}
\end{lem}
\noindent This semantics is very similar to its late counterpart. The reason for this is that in the weak late operational semantics, the instantiations of input bound names occur inside the transition before the succeeding $\tau$-chain. This becomes apparent when we compare the lifted rules for $\mathbf{Input}$. In the late semantics, it looks like an early transition since it contains the name $u$ received in the input. The rules $\mathbf{Close}$ and $\mathbf{Comm}$ also behave in the same way. We have proven Lemmas \ref{earlyLateOutput}, \ref{earlyLateBoundOutput}, \ref{earlyLateInput}, \ref{lateEarlyInput} and Theorem \ref{earlyLateTau} for the correspondence of the weak late and early transition systems.

We do encounter the same problem when trying to lift the $\Gta{\widehat\alpha}_e$ transitions in that $\mathbf{Match}$, $\mathbf{Mismatch}$, $\mathbf{Sum}$ and $\mathbf{Replication}$ cannot be lifted, for the same reason as in the late semantics. The lifted early rules correspond more closely to their strong counterparts than the lifted late rules correspond to theirs. The weak early and late rules are very similar to each other since the $\mbox{Input}$-rules behave in the same intuitive manner.  The difference between the two semantics is not so much in the operational rules as in the definition of simulation.\\
\begin{defi}
\label{weakEarlySimulation}
The agent $P$ can weakly early simulate the agent $Q$ preserving $\mathcal{R}$, written $P\weakSim{e\mathcal{R}}Q$, if
\[
(\forall a~x~Q'.~Q\gta{\bar{a}(x)}_eQ' \land x~\sharp~P \Longrightarrow\exists P'.~P\Gta{\bar{a}(x)}_eP' \land (P',~Q') \in \mathcal{R})~\land
\]
\[
(\forall \alpha~Q'.~Q\gta{\alpha}_eQ' \Longrightarrow \exists P'.~P\Gta{\widehat{\alpha}}_eP' \land (P',~Q') \in \mathcal{R})
\]

\end{defi}

\begin{defi}Weak early bisimulation equivalence, $\approx_e$, is the largest relation satisfying:
\label{weakEarlyBisimulation}
\[
  P \weakBisim_e Q \define P\weakSim{e\weakBisim_e}Q\mathrel{\land}Q\weakSim{e\weakBisim_e}P
\]
\end{defi}

\begin{thm}
$\approx$ is preserved by all operators except $+$ and input-prefix.
\end{thm}
\begin{proof}Similar to the proof for Theorem \ref{weakBisimPres}.\end{proof}
\noindent Weak early bisimulation is not a congruence for the same reason as weak late bisimulation, and in order to create a congruence we need to define a weak early congruence simulation, $\weakCongSim_e$, by replacing the $\Gta{\widehat\alpha}_e$ in Def.\ \ref{weakEarlySimulation} by $\Gta{\alpha}_e$.

We can now define our weak early congruence.
\begin{defi}
\label{weakEarlyEquivalence}
$P \cong_e Q \define P\; \weakCongSim_{e\approx_e}\; Q\; \land\; Q\; \weakCongSim_{e\approx_e}\; P$
\end{defi}

\begin{thm}
$\cong_e$ is preserved by all operators except input-prefix.
\end{thm}
\begin{proof}
Similar to the proof for Theorem \ref{weakEqPres}.
\end{proof}
\begin{lem}
$\cong_e^s$ is a congruence.
\end{lem}
\begin{proof}
Proved in a similar way as Lemma \ref{weakEqCong}.
\end{proof}
\subsection{Relationships between equivalences}
Not surprisingly, strong early and weak early relations enjoy the same inclusion properties as their late counterparts, i.e. $\bisim_e\; \mathrel{\subseteq}\; \weakCong_e\; \mathrel{\subseteq}\; \weakBisim_e$. Furthermore, $\sim\; \mathrel{\subseteq}\; \sim_e$.

The proof for the latter is more involved and requires correspondance proofs between strong early and late actions. The connection we have proved between them is that every early $\tau$-transition has a corresponding late $\tau$-transition and vice versa. More precisely, the following lemmas are proven:
\begin{lem}
$P\gta{\bar{a}b}_e P'$ iff $P\gta{\bar{a}b} P'$
\label{earlyLateOutput}
\end{lem}
\begin{proof}
By induction over the possible output transitions.
\end{proof}
\begin{lem}
$P\gta{\bar{a}(x)}_e P'$ iff $P\gta{\bar{a}(x)} P'$
\label{earlyLateBoundOutput}
\end{lem}
\begin{proof}
By induction over the possible bound output transitions.\\
Before induction, the transitions are $\alpha$-converted such that $x$ is fresh for $a$ and $P$. In the $\mbox{Open}$ cases, Lemma \ref{earlyLateOutput} is used.
\end{proof}
\begin{lem}
If $P\gta{a(x)} P'$ then $P\gta{au}_e P'\{u/x\}$
\label{earlyLateInput}
\end{lem}
\begin{proof}
By induction over the possible input transitions. Before induction, the late transition is $\alpha$-converted such that $x$ is fresh for $a$, $u$ and $P$.
\end{proof}
\begin{lem}
If $P\gta{au}_e P'$ then for all name contexts $\mathcal{C}$, there exists an $x$ and a $P''$ s.t. $P\gta{a(x)} P''$, $P' = P''\{u/x\}$ and $x\fresh\mathcal{C}$\\
\label{lateEarlyInput}
\end{lem}
\begin{proof}By induction over the possible input-transitions. When doing the induction, the last conjunct of the goal is not used but only the first two ones. We can then take the results from the induction and eliminate the existential quantifiers, pick a new fresh name $x'$ which is fresh for $P''$ and $\mathcal{C}$ and instantiate the goal with $x'$ and $(x~x')\bullet P''$.
\end{proof}
\noindent We can now prove our theorem.
\begin{thm}
$P\gta{\tau}_e P'$ iff $P\gta{\tau} P'$
\label{earlyLateTau}
\end{thm}
\begin{proof}By induction over the possible $\tau$-transitions. In the $\mathbf{Open}$, $\mathbf{Comm}$ and $\mathbf{Close}$ cases, Lemma \ref{earlyLateOutput}, \ref{earlyLateBoundOutput}, \ref{earlyLateInput} and \ref{lateEarlyInput} are used.
\end{proof}
\noindent We can now continue with our correspondence proofs between late and early semantics.
\begin{lem}
If $P\leadsto_{\mathcal{R}}Q$ then $P\leadsto_{e\mathcal{R}} Q$
\label{earlyLateSim}
\end{lem}
\begin{proof}By case analysis of Def.\ \ref{earlySimulation}. Lemmas \ref{earlyLateOutput}, \ref{earlyLateBoundOutput}, \ref{earlyLateInput}, \ref{lateEarlyInput} and Theorem \ref{earlyLateTau} are used to transform the early simulations to late ones and back again after applying Def.\ \ref{simulation}. Lemma \ref{lateEarlyInput} has the context $\mathcal{C}$ instantiated as $P$ to ensure that the generated bound name is fresh for $P$ as required by Def.\ \ref{simulation}.
\end{proof}
\noindent We now prove that all late bisimilar processes are also early bisimilar.
\begin{thm}
If $P\sim Q$ then $P\sim_e Q$
\label{lateEarlyBisim}
\end{thm}
\begin{proof}By coinduction using Prop.\ \ref{bisimWeakCoinduct} and setting $\mathcal{X}$ to $\sim$. Lemma \ref{earlyLateSim} then proves our goal.
\end{proof}
\begin{cor}
If $P\sim^sQ$ then $P\sim_e^sQ$
\label{lateEarlyEq}
\end{cor}
\begin{proof}
Follows trivially from Theorem \ref{lateEarlyBisim}.
\end{proof}
\noindent With these we can very easily prove our theorems about structural congruence for early.
\begin{cor}
\[
\begin{array}[t]{r@{\; }l}
\mbox{If}\; P\equiv Q\; \mbox{then}& P\sim_e Q\\
\mbox{and} & P\sim_e^s Q
\end{array}
\]
\end{cor}
\begin{proof}
Follows trivially from Theorems \ref{congBisim}, \ref{lateEarlyBisim}, \ref{congEq} and Cor.\ \ref{lateEarlyEq}.
\end{proof}
\noindent  Finally, for the weak early semantics:
\begin{cor}
\label{weakEarlyCor}
\[
\begin{array}[t]{l}
\mbox{If}\; P\bisim_e Q\; \mbox{then}\; P\weakCong_e Q\\
\mbox{If}\; P\bisim_e^s Q\; \mbox{then}\; P\weakCong_e^s Q\\
\mbox{If}\; P\weakCong_e Q\; \mbox{then}\; P\weakBisim_e Q\\
\end{array}
\]
\end{cor}
\begin{proof}
Similar to their corresponding proofs in section \ref{weakLateStructCong}.
\end{proof}
\noindent From this our structural congruence results follow trivially.
\begin{cor}
\[
\begin{array}[t]{r@{\; }l}
\mbox{If}\; P\equiv Q\; \mbox{then}& P\weakCong_e Q\\
\mbox{and}& P\weakBisim_e Q\\
\mbox{and}& P\weakCong_e^s Q\\
\mbox{and}& P\weakBisim_e^s Q
\end{array}
\]
\end{cor}
\begin{proof}
From Theorems \ref{congBisim}, \ref{congEq} and Cor.\ \ref{weakEarlyCor}.
\end{proof}

\section{Results and Conclusions}

\subsection{Current Status}
We have used the new nominal datatype package in Isabelle to model the
$\pi$-calculus and our results are very encouraging. We have proved a
substantial part of \cite{parrow:acomp}, in particular preservation properties of
strong and weak
bisimulation for both late and early operational semantics. Other results include that all late $\tau$-transitions have a corresponding early one and vice versa and that all late bisimulation relations have an early counterpart. Moreover, we have proven that all the bisimulation relations we have investigated contain structural congruence. We have created a
substantial library concerning the fundamental mechanisms in the
$\pi$-calculus, such as substitution and transitions. One of our main
contributions is that the proofs resemble the ones on paper very
closely, since we make precise the traditional ``hand waving'' with
respect to bound names.  Since we are using Isabelle, we can write our
proofs in a very readable form using \emph{Isar} \cite{wenzel99isar}. We believe this to be the most extensive  formalisation of a process calculus ever done inside a theorem prover.

In recent work we put our formalisation to the test by proving that the axiomatisation of strong late bisimilarity is sound and complete \cite{bengtson:sos2007}. The proofs were complex, but again mapped their pen-and-paper equivalents very closely and we made extensive use of the foundation provided in this paper.

The nominal package is still work in progress and it is constantly
being updated. One recent addition allows for users to define functions on their nominal datatypes using an automatically generated recursion combinator \cite{urban:rcnd}. At the moment the only function we use is substitution.
\subsection{Related Work}
The $\pi$-calculus has been subjected to many attempts at formalisations. Gabbay made a formalisation in \cite{GabbayMJ:picfm} utilising FM set theory, the precursor of nominal logic. His work is mathematically close to ours. The rest of this section will focus mainly on formalisations which have been subject to mechanisation inside a theorem prover.
Early sketches in HOL include \cite{694551,melham94mechanized}. Later attempts have also been made using de-Bruijn indices where names are encoded using natural numbers.
The most extensively used approach is higher order abstract syntax (HOAS) where weak HOAS is the technique most similar to ours. We here comment on the more important approaches.

de Bruijn indices are heavily used in software which reasons about
terms with binders; an example for the $\pi$-calculus is  the Mobility Workbench
\cite{victor:mwb}. They work well in these environments as they have
very nice algorithmic properties. However, these properties do not provide an intuitive mathematical framework. In \cite{hirschkoff:affopt}, Daniel Hirschkoff formalised a subset of the $\pi$-calculus excluding sum, match and mismatch in Coq using de Bruijn indices. The theories formalised was that early bisimulation is a congruence as well as the structural congruence results. Preliminary work was also made to help formalise Milner's encoding of the $\lambda$-calculus \cite{milner:fap}. Hirschkoff writes the following:
\begin{quote}
Technical work, however, still represents the biggest part of our implementation, mainly due to the managing of de Bruijn indexes\ldots Of our 800 proved lemmas, about 600 are concerned with operators on free names.
\end{quote}

Fraenkel Mostowski set theory was one of the first serious attempts to fomalise nominal logic. It is standard ZF set theory but with an extra freshness axiom added. In \cite{GabbayMJ:picfm}, Gabbay formalises a portion of the {\picalc} in FM. In this approach a {\new}-quantifier (new quantifier) is used to generate names which are fresh for the current context. The nominal package does not provide support for this quantifier, but the same effect is achieved by instantiating our rules with a set of context names. Gabbay also started work on incorporating a framework for FM inside Isabelle \cite{GabbayMJ:autfms} with which formalisations such as ours could be made. Unfortunately, this early version of nominal logic was incompatible with the axiom of choice and had to be used in Isabelle/PURE -- a bare boned set of theories. This choice of framework was necessary since Isabelle/HOL contains the axiom of choice, but the attempt was later abandoned.

HOAS has been used to model
the $\pi$-calculus in both Coq \cite{honsell:picalculus}, by Honsell
et. al., and in Isabelle by Röckl and Hirschkoff
\cite{roeckl:sepi}. In \cite{honsell:picalculus} the late operational semantics is encoded together with late strong bisimulation. The proved results include that the algebraic laws presented in \cite{parrow:acomp} are sound where the non-trivial proofs include preservation results for bisimulation and the results for structural congruence. When using HOAS terms, binders are represented as
functions of type \texttt{name->term}. However, if these functions
range over the entire function space they may produce exotic terms, so
the formalisations need to ensure that those are avoided. In
\cite{roeckl:sepi}, a special well-formedness predicate is used to
filter out the exotic terms. Another problem is that since abstraction is handled by the meta-logic of the theorem prover, reasoning about binders at the object level can become problematic. In \cite{honsell:picalculus} we can read:
\begin{quote}
The main drawback in HOAS is the difficulty of dealing with
metatheoretic issues concerning names in process contexts, \emph{i.e.}
terms of type \texttt{name->proc}. As a consequence, some
metatheoretic properties involving substitution and freshness of names
inside proofs and processes, cannot be proved inside the framework and
instead have to be postulated. 
\end{quote}
Our approach is completely free from any extra axioms, and
since nominal logic is a first order approach we do not have exotic terms. Moreover, freshness conditions are part of the nominal infrastructure and all such conditions are explicitly known at the object level and do not have to be postulated, thus no extra infrastructure for choosing particular names is needed. 
\subsection{Impact and Further Work}
Theorem provers suffer from a somewhat well-deserved reputation of
being hard to use for the uninitiated. However, having theories
formalised by a computer has significant advantages and making theorem
provers easy to use for the general engineer is a high
priority. We believe that our work helps in this venture. The
challenging part has been to create inductive rules and easy-to-use definitions for simulation and
bisimulation. With this done the actual proofs done in the theorem
prover are not much harder than the ones done on paper.

Our next goal will be to provide support for model- and bisimulation
checking on actual protocols such as ad-hoc routing. Particularly
processes with infinite state space are of interest as these cannot be
handled by automatic tools like the Mobility Workbench.

There are several variants of the $\pi$-calculus, polyadic
$\pi$-calculus and higher order $\pi$-calculus just to name two. We
believe that our definitions for simulation and bisimulation can
easily be transfered to many other calculi.

\subsubsection*{Acknowledgments}

We would like to thank Stefan Berghofer for his generous help with the inner workings of Isabelle, Christian Urban for developing the nominal datatype package and providing extensive support and insights, and Lars-Henrik Eriksson for discussions on theorem provers. We would also like to thank the anonymous referees for their many helpful and constructive comments.


\bibliographystyle{plain}
\bibliography{bibliography}

\end{document}